\documentclass{aa}
\usepackage{graphicx,epsfig}

\newcommand{\ergscm}{erg\,s$^{-1}$\,cm$^{-2}$}
\newcommand{\ergscma}{erg\,s$^{-1}$\,cm$^{-2}$\,\AA$^{-1}$}
\begin{document}
   \title{Spectral line variability amplitudes in AGNs  
          \thanks{Based on observations taken at the German-Spanish
 Astronomical Center Calar Alto} $\!^{,}$
          \thanks{Based on observations obtained with the Hobby-Eberly
               Telescope, which is a joint project of the University
               of Texas at Austin, Pennsylvania State
           University, Stanford University, Ludwig-Maximilians-Universit\"at
            M\"unchen, and Georg-August-Universit\"at G\"ottingen.}
 }
   \author{W. Kollatschny$^1$, M. Zetzl$^1$, and M. Dietrich$^2$}

   \offprints{W. Kollatschny}

   \institute{$^1$ Institut f\"{u}r Astrophysik, Universit\"{a}t G\"{o}ttingen,
              Friedrich-Hund-Platz 1, D-37077 G\"{o}ttingen, Germany\\
              \email{wkollat@astro.physik.uni-goettingen.de}\\
              $^2$ Department of Astronomy, The Ohio State University,
              4055 McPherson Lab, 140 W. 18th Ave., Columbus, OH 43210, USA}

   \date{Received date, 2005; accepted date, 2005}

   \abstract{
We present the results of a long-term variability campaign of very broad-line
AGNs with line widths broader than FWHM $>$ 5000~kms$^{-1}$.
The main goal of our investigation was to study whether the widths of the 
optical broad emission lines are correlated with the optical intensity 
variations on timescales of years. Our AGN sample consisted of 10 objects.
We detected a significant correlation between optical continuum variability 
amplitudes and  H$\beta$ emission line widths (FWHM) and, to a lesser degree, 
between  H$\beta$ line intensity variations and H$\beta$ equivalent widths.
We add the spectroscopic data of variable AGNs from the literature to
supplement our sample. 
The AGNs from other optical variability campaigns with different line-widths
helped to improve the statistical significance of our very broad-line AGN
sample. After including the data on 35 additional galaxies, the
correlation between 
optical continuum variability amplitudes and H$\beta$ emission line widths
becomes even more significant and the probability that this is a random
correlation drops to 0.7 percent.
   }
  \authorrunning{W. Kollatschny}
  \titlerunning{Spectral variability amplitudes}

  \maketitle
%

\section{Introduction}

Seyfert\,1 galaxies and quasars vary 
on timescales of days to years in the optical 
(e.g. Peterson et al. \cite{peterson04}, Kaspi et al.\cite{kaspi00}).
The variablity amplitudes of their continuum fluxes range from a few
percent to more than a hundred percent (e.g. de Vries et al. \cite{devries05}).
Variations in the continuum from the radio to the gamma-ray range have 
different timescales (e.g. Peterson et al. \cite{peterson00}),
which is a situation that has to be taken into account in variability studies.
This makes it necessary to examine only the individual frequency ranges.
Even though considerable
effort has been made over the last years, several details of
the physical processes leading to variability are still unknown 
(Peterson \cite{peterson01}, Blandford 
\cite{blandford04}, Horne et al. \cite{horne04} and references therein).

In an earlier extensive investigation, Giveon et al. (\cite{giveon99})
studied a sample
of PG quasars with line widths in the range of full width at half maximum
(FWHM) $\sim 2000$ to 5000 km\,s$^{-1}$. They searched for correlations between
the measured variability properties and other parameters of the
studied quasars. However, they did not look explicitly for correlations
between the emission line width and other intrinsic parameters.

In this paper we focus on optical continuum variations and emission-line 
intensity variations in AGNs over timescales of years, which is comparable to 
the expected dynamical time scale of the broad-line region (BLR). 
Observed variations in the emission lines in the spectra of AGNs are
caused by variations in the central ionizing source. In this paper
 we investigate whether there are correlations between the variability 
amplitudes in the continuum and properties of the spectral lines
to gather information about the structure of the BLR.

Nearly nothing is known about the optical variability behavior of very
broad-line AGNs because most studies have so far been concentrated
on AGNs with 
FWHM $\sim 2000$ to 5000 km\,s$^{-1}$. However, one of the most variable 
Seyfert galaxies known so far is NGC\,7603. This galaxy with emission-line 
widths of FWHM\,$\simeq$\,6500 km\,s$^{-1}$ varied by a factor of 5 to 10 in 
the optical continuum and in the Balmer lines over a period of 20 years 
(Kollatschny et al. \cite{kollatschny00}). 
Hence, we selected Seyfert galaxies with emission-line 
widths of more than 5000 km\,s$^{-1}$. We investigate the variability 
behavior of a sample of ten Seyfert galaxies. In addition, we compare 
our results with results from earlier campaigns
investigating AGNs with
narrower emission-line widths ($\sim 2000$ to 5000 km\,s$^{-1}$).

\section{Observations}
\subsection{The spectroscopic variability sample}
The broad emission lines in AGNs have typical line widths of more than 
2000~kms$^{-1}$ (Hao et al. \cite{hao05}). The so-called narrow-line Seyfert 1 galaxies 
(FWHM\,$\leq$ 2000 km\,s$^{-1}$, e.g. Osterbrock, D., \& Pogge, 
\cite{osterbrock85}) have been the subject of many papers in previous years
(e.g.  Williams et al. \cite{Williams02}, Romano et al. \cite{Romano04}).
On the other hand, 15 to 20 percent of all AGNs have line widths
that are broader than 
5000~kms$^{-1}$, as seen in a line-width distribution of AGNs recently
published by Hao et al. (\cite{hao05}), who analyzed
the AGN spectra of the Sloan Digital Sky Survey.

Most of the variability studies of AGNs have so far been made for galaxies
with line widths of less than 6000~kms$^{-1}$. 
Nearly nothing is known about the long-term variability statistics of
the spectral lines of very broad-line AGNs
- except for 3C390.3 (e.g. Veilleux \& Zheng \cite{veilleux91},  
                            Dietrich et al.  \cite{dietrich98},
                            O'Brien et al. \cite{OBrien98}).

We started a campaign of observing the
 long-term spectral variability of the very broad-line 
AGN class more than 10 years ago. We selected as many as possible bright 
($m_{V} <$ 16), nearby northern AGNs of this very broad-line class from the 
literature
(e.g. Osterbrock et al. \cite{osterbrock76}, 
Apparao et al. \cite{apparao78}, Smith \cite{smith80}, Bond \cite{bond77}). 
Our main goal was to study the emission-line variability of very broad line 
AGNs with those telescopes in the 2\,m to 4\,m class that were available
at this time.
We had to include objects with higher redshifts (z$>$0.05) because very 
broad-line AGNs are rare in the local universe.
Our target AGNs include intrinsically luminous, as well as faint, AGNs.

Finally, a sample of ten AGNs was compiled, all having H$\beta$ emission-line 
widths of FWHM\,$\simeq 5000$ to 13\,000 km\,s$^{-1}$ (Table 1).
For comparison we investigated two narrow-line Seyfert 1 galaxies
with line-widths of less than 2000 kms$^{-1}$) (Mrk~110, Ton~256).

The galaxies in our sample, their mean H$\beta$ line widths,
 their redshifts, as well as
 their apparent and absolute brightness
(taken from the catalogue of V\'eron \& V\'eron \cite{Veron01})
are given in Table~1.
In addition, the  [\ion{O}{iii}]$\lambda$5007 line widths,
the [OIII]/H$\beta$ line ratios derived from our spectra,
the radio power, the references from which the radio data were selected,
and the radio optical spectral indices $\alpha_{\mathrm{RO}}$ are listed. 
The radio optical spectral index $\alpha_{\mathrm{RO}}$ has been calculated by
us using the formula

\begin{equation}
  \label{eq:alpha_ro}
\alpha_{RO} = - \frac{\log{(L_{R}/L_{O})}}{\log{(\nu_{R}/\nu_{O})}}~.
\end{equation}

\begin{table*}
\caption{AGN broad-line variability sample}
\tabcolsep3.2mm
\begin{tabular}{lccccccrrcc}
\hline
\noalign{\smallskip}
Object & FWHM H$\beta$ & z & m$_{v}$ & M$_{B}$& FWHM [OIII]& [OIII]/H$\beta$  & $\log{P_{5}}$ &$\alpha_{\mathrm{RO}}$ & Ref. \\
 &  [kms$^{-1}]$       &           &  &  & [kms$^{-1}$] &  & [W\,Hz$^{-1}$]   &   \\
(1) & (2) & (3) & (4) & (5) & (6) &(7) & (8) & (9) & (10)  \\
\noalign{\smallskip}
\hline
\noalign{\smallskip}
3C390.3    & 11530. & 0.056     &  15.4 & -21.6& 799   & 0.57 & 24.659 &  0.43 & 1 \\
4U0241+61  &  8720. & 0.044     &  12.2 & -25.0& 814   & 0.32 & 28.607 &  0.91 & 2 \\
E1821+643  &  6230. & 0.297     &  14.2 & -27.1& 854   & 0.29 & 24.755 &  0.04 & 3 \\
Mrk~876    &  7390. & 0.129     &  15.5 & -23.5& 918   & 0.20 & 22.277 & -0.15 & 1 \\
Mrk~926    &  8510. & 0.047     &  13.8 & -23.1& 766   & 0.78 & 20.754 & -0.40 & 1 \\
NGC~6814   &  7560. & 0.005     &  14.2 & -17.5& 757   & 2.51 & 19.374 & -0.24 & 1 \\
OX~169     &  5010. & 0.211     &  15.7 & -24.7& 1053  & 0.17 & 25.288 &  0.32 & 1 \\
PKS2349-14 & 5130.  & 0.174     &  15.3 & -24.7& 719   & 0.31 & 24.917 &  0.25 & 1 \\
NGC~7603   &  6560. & 0.0295    &  14.0 & -21.5& 785   & 0.15 & 21.512 & -0.14 & 1 \\
V396HER    & 12330. & 0.175     &  16.4 & -23.3& 764   & 0.21 & 24.676 &  0.31 & 2 \\
\noalign{\smallskip}                                     
\hline                                                                 
\noalign{\smallskip}                                     
Mrk~110    &  1670. & 0.0355    &  15.4 & -20.6&596    & 0.54 & 21.242 & -0.13 & 1 \\
Ton256     &  1460. & 0.131     &  15.4 & -23.5& 635   & 1.25 & 22.514 & -0.11 & 1 \\
\noalign{\smallskip}
\hline
\noalign{\smallskip}
\end{tabular}
\\
Ref. 1: Xu et al. (\cite{xu99})\\
Ref. 2: Gregory et al. (\cite{gregory91})\\
Ref. 3: Kolman et al. (\cite{kolman91})\\
\end{table*}

\subsection{Observations and data reduction}

We took optical spectra of our galaxies over a period of 3 to 7 years. Details
of the observations, such as the observing dates,
the corresponding Julian dates, 
the telescopes we used, the wavelength coverage of the spectra, and exposure 
times, are listed in Table~2.
In most cases we observed one spectrum per galaxy per year.

\begin{table}
\caption{Log of observations}
\tabcolsep+0.1mm
\begin{tabular}[H]{ccccr}
\hline
\noalign{\smallskip}
Julian Date & UT Date & Telescope & $\lambda\lambda$ & Exp. time \\
2\,400\,000+&         &           &    [\AA]           & [sec.]   \\
(1) & (2) & (3) & (4) & (5) \\
\noalign{\smallskip}
\hline
\noalign{\smallskip}
3C~390.3 \\
\hline
      48089            & 1990-07-17 & CA 3.5   & 3970  -- 7300             & 3600\\
      48475            & 1991-08-06 & CA 2.2   & 3360  -- 8290             & 3300\\
      48866            & 1992-09-01 & CA 2.2   & 3850  -- 9150             & 3600 \\
      49595            & 1994-08-30 & CA 2.2   & 4520  -- 8530             & 3600 \\
\hline                                                               
4U~0241+61\\                                                       
\hline                                                               
      48475            & 1991-08-07 & CA 2.2   & 3360  -- 8290             & 1800 \\
      48861            & 1992-08-27 & CA 2.2   & 3850  -- 9150             & 3600 \\
      49596            & 1994-09-01 & CA 2.2   & 4520  -- 8530             & 3600 \\
\hline                                                               
E~1821+643\\                                                        
\hline                                                               
      48095            & 1990-07-23 & CA 3.5   & 6050  -- 9400             & 5280   \\
      48476            & 1991-08-07 & CA 2.2   & 5130  -- 10800            & 3600   \\
      48813            & 1992-07-09 & CA 3.5   & 3670  -- 9690             & 1200   \\           
      48860            & 1992-08-25 & CA 2.2   & 3860  -- 9150             & 3600   \\           
      49597            & 1994-09-01 & CA 2.2   & 5960  -- 10000            & 3300   \\           
\hline                                                               
 MRK~876 \\                                                           
\hline                                                               
      48093            & 1990-07-20 & CA 3.5   & 4970  -- 8310             & 6000  \\
      48474            & 1991-08-05 & CA 2.2   & 3350  -- 8300             & 4500  \\
      48814            & 1992-07-10 & CA 3.5   & 3670  -- 9690             & 1200  \\
      48862            & 1992-08-27 & CA 2.2   & 3860  -- 9150             & 3000  \\
      49598            & 1994-09-02 & CA 2.2   & 4980  -- 9000             & 2400  \\
\hline                                                               
MRK~926\\                                                              
\hline                                                               
      48090            & 1990-07-18 & CA 3.5   & 3970  -- 7300             & 1500 \\                                          
      48092            & 1990-07-20 & CA 3.5   & 6450  -- 7320             & 4000 \\                                          
      48478            & 1991-08-10 & CA 2.2   & 4070  -- 5560             & 3600 \\                                  
      48480            & 1991-08-11 & CA 2.2   & 5510  -- 7630             & 3000 \\                                  
      48816            & 1992-07-13 & CA 3.5   & 3670  -- 5900             & 2400 \\                               
      49215            & 1993-08-16 & CA 2.2   & 3750  -- 8170             & 1800 \\ 
      49577            & 1994-08-12 & CA 2.2   & 3800  -- 9670             & 900 \\ 
      49596            & 1994-08-31 & CA 2.2   & 4520  -- 8530             & 3000 \\                                          
      50395            & 1996-11-07 & ESO 2.2   & 3810  -- 7980             & 2400 \\ 
      50635            & 1997-07-05 & ESO 2.2   & 4100  -- 7470             & 720 \\ 
                                                                           
\hline                                                               
NGC~6814\\                                                            
\hline                                                               
      48812            & 1992-07-09 & CA 3.5   & 3660  -- 9690             & 6300  \\
      48866            & 1992-08-31 & CA 2.2   & 3850  -- 9150             & 1800   \\
\hline                                                               
OX~169 \\                                                               
\hline                                                               
      48093            & 1990-07-21 & CA 3.5   & 4970  -- 8310             & 5400 \\
      48865            & 1992-08-30 & CA 2.2   & 3850  -- 9150             & 3600 \\
      49598            & 1994-09-02 & CA 2.2   & 4980  -- 9000             & 5400 \\
\hline                                                               
PKS~2349-014 \\                                                   
\hline                                                               
      48094            & 1990-07-22 & CA 3.5   & 4970  -- 8290             & 6000 \\
      48864            & 1992-08-30 & CA 2.2   & 3860  -- 9160             & 3600 \\
      49598            & 1994-09-03 & CA 2.2   & 4980  -- 9000             & 3600 \\
\hline                                                               
V396~Her  \\                                                          
\hline                                                               
      48093            & 1990-07-21 & CA 3.5   & 4970  -- 8290             & 3600 \\
      48864            & 1992-08-29 & CA 2.2   & 3860  -- 9160             & 3600 \\
      49596            & 1994-08-31 & CA 2.2   & 4520  -- 8530             & 5450\\
\hline                                                               
TON~256 \\                                                             
\hline                                                               
      48094            & 1990-07-21 & CA 3.5   & 4970  -- 8310             & 7200\\
      48475            & 1991-08-06 & CA 2.2   & 3340  -- 8290             & 4560\\
\hline
\end{tabular}
\end{table}

The spectra were obtained at Calar Alto Observatory in Spain with the 2.2\,m 
and 3.5\,m telescopes. Individual exposure times range from 15 minutes to 2 
hours (see Table~2).
In most cases a Boller \& Chivens spectrograph was attached to the
telescope, but several observations were recorded using the TWIN spectrograph
at the 3.5~m telescope. The spectrograph slits had projected widths 
of 2 to 2.5 arcsec and 2 arcmin length.
Typical seeing conditions were 1 to 2 arcsec.
The slit was oriented in the north-south direction, in most cases,
to minimize the impact of the light loss caused by differential refraction.
We used different CCD detectors in the course of our monitoring program. 
Details of the CCD detectors we used are given
in Table~3.

 \begin{table}[tbp]
  \caption{CCDs - technical data}
    \centering
       \leavevmode
\tabcolsep1.1mm
         \begin{tabular}{lllll}
 \hline
Date &CCD & Dimension [Pixel] & \multicolumn{2}{l}{Pixel Size}   \\
     &&                   & \multicolumn{1}{c}{[$\mu{}$m]} &  \multicolumn{1}{c}{[$''$]} \\ 
 \hline
July      1990 & RCA \#11             &  1024 $\times$ 640     & 15    &0.58  \\ 
August    1991 & GEC \#13             &  1155 $\times$ 768     & 22.5  &1.32  \\
July      1992 & RCA \#10             &  1024 $\times$ 640     & 15    &0.56  \\
July      1992 & GEC \#14             &  1155 $\times$ 768     & 22.5  &0.84  \\
August    1992 & TEK \#6              &  1024 $\times$ 1024    & 24    &1.41  \\
September 1994 & TEK \#13             &  1024 $\times$ 1024    & 24    &1.41  \\
November  1996 & LORAL                &  2048 $\times$ 2048    & 15    &0.26  \\
July      1997 & LORAL                &  2048 $\times$ 2048    & 15    &0.26  \\
 \hline
  \end{tabular}
     \label{tab:techn.CCD}
          \end{table} 
In most cases, our spectra cover a wavelength range from 3800\,\AA\ to 
7400\,\AA\ , typically (see Table~2) with
a spectral resolution of 3 to 7\,\AA\ per pixel.
HeAr spectra were taken after each object exposure for
wavelength calibration. Various standard stars were
observed for flux calibration. 
One spectrum for PKS~2349-14 was obtained with the 9.2m 
Hobby-Eberly Telescope (HET) at McDonald Observatory on August 24, 2004.
We used the Marcario Low Resolution Spectrograph and the exposure time was
20 minutes. Details of the observing conditions with the HET can be found
e.g. in Kollatschny et al. (\cite{kollatschny01}). 
 
The reduction of the spectra (bias subtraction, cosmic ray correction,
flat-field correction, 2D-wavelength calibration, night sky subtraction,
flux calibration) was done in a homogeneous way with IRAF reduction
packages \footnote{IRAF is distributed by the National Optical Astronomy
Observatories}. We extracted spectra of the central 5 arcsec.
The spectra have signal-to-noise ratios (S/N) of 20 to 40 in the continuum.
Great care was taken to achieve a very accurate relative intensity
calibration.
All spectra were rebinned to the same spectral resolution and calibrated to 
the same absolute [\ion{O}{iii}]$\lambda$5007 flux for each corresponding 
galaxy. Our absolute [\ion{O}{iii}]$\lambda$5007 flux was obtained from those
spectra that were taken under photometric conditions. 
In addition, the accuracy of our [\ion{O}{iii}]$\lambda$5007 flux 
calibration was tested for all forbidden emission-lines over the entire 
spectrum. We calculated difference spectra for all our epochs with respect to 
the mean spectrum of our variability campaign. Thus we minimized the
contamination by the host galaxy.
The stellar Mg~I~b absorption feature at 5175\,\AA\ is very weak
in all our AGN spectra except for NGC~6814
(see Figs.~\ref{fig:lk_1},\ref{fig:lk_2}).
Therefore, the relative contribution of the host galaxy flux    
to the ionizing continuum flux at 5100\,\AA\
should be less than 20 percent in our spectra.
Corrections for small spectral 
shifts ($<$ 0.5 \AA ) and for different scaling factors were executed by 
minimizing the residuals of the narrow emission lines in the difference 
spectra. 
We achieved a relative accuracy 
of 3 to 5 percent by means of the [\ion{O}{iii}]$\lambda$5007 flux  
calibration.

\section{Results}
\begin{figure}[hbp]
        \centering
        \includegraphics[width=8.5cm]{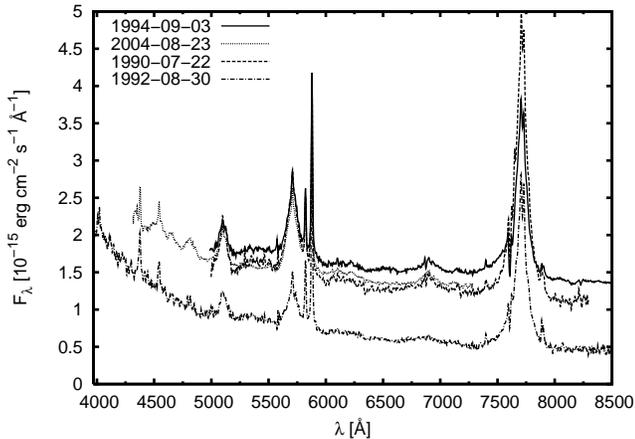}
        \caption{Spectra of the broad-line AGN PKS~2349-14 taken at Calar Alto
 Observatory between 1990 and 1994. A spectrum 
taken with the 9.2m Hobby-Eberly Telescope (HET) in 2004 is shown 
for comparison.}
        \label{fig:pks2349_alle_c}
\end{figure}

In Fig.~\ref{fig:pks2349_alle_c} the spectra of PKS\,2349$-$14 are
shown to illustrate the range
of spectral variations for the AGNs in our study.
Between 1990 and 1994, there was a sharp drop by more than 50 percent
in the continuum flux, as well as in the broad emission-line intensities.
 A spectrum of PKS~2349-14 taken with the 
9.2m Hobby-Eberly Telescope (HET) in 2004 is shown for comparison in Fig.~1.
It appears that PKS\,2349$-$14 as observed in 2004 has recovered the 
state that was measured in 1990.

We measured the rest frame continuum flux at 5100\,\AA\ and the integrated 
line intensities of the broad emission lines in the relative flux calibrated 
spectra of our sample.
The wavelength boundaries we used for our continuum flux
measurements at 5100\,\AA\
are given in Table~4.
The underlying continuum of the H$\beta$ and H$\alpha$ emission lines
was interpolated between local minima on the short and
long-wavelength sides of the lines (pseudo-continua). We integrated
the total emission line fluxes above the continuum
between the wavelength boundaries given in Table~4.

\begin{table}
\caption{Extraction windows of the continuum and emission-line flux
integration limits.}
\tabcolsep+1.2mm
\begin{tabular}{lccc}
\hline
&\multicolumn{3}{c}{Wavelength range [\AA{}]}\\
&\multicolumn{3}{c}{\hrulefill}\\
\noalign{\smallskip}
Object    & Continuum & H$\beta{}$ & Pseudo-continuum\\
          &           & H$\alpha{}$ & \\
\noalign{\smallskip}
 (1)      &(2)        &(3)    & (4) \\
\noalign{\smallskip}
 \hline
3C~390.3   & 5368 -- 5398  & 4968 -- 5318               & 4800 -- 5410\\
          &               & 6628 -- 7188               & 6552 -- 7280\\
4U~0241+61 & 5268 -- 5288  & 4998 -- 5264               & 4800 -- 5288\\
          &               & 6598 -- 7198               & 6400 -- 7300\\
E~1821+643 & 6626 -- 6666  & 6200 -- 6550               & 6150 -- 6666\\
          &               & 8098 -- 8878               & 8050 -- 9100\\
Mrk~876    & 5740 -- 5780  & 5362 -- 5686               & 5330 -- 5780\\ 
          &               & 7030 -- 7558               & 6900 -- 7880\\
Mrk~926    & 5335 -- 5350  & 4978 -- 5378               & 4750 -- 5350\\      
          &               & 6700 -- 7100               & 6472 -- 7280 \\       
NGC~6814   & 5112 -- 5152  & 4798 -- 4962               & 4650 -- 5152\\
          &               & 6310 -- 6706               & 5950 -- 6850\\
OX~169     & 6152 -- 6192  & 5798 -- 6108               & 5740 -- 6192\\ 
          &               & 7742 -- 8112               & 7625 -- 8270\\
PKS~2349-14& 5966 -- 6026  & 7346 -- 7958               & 7240 -- 8250\\
          &               & 7346 -- 7958               & 7240 -- 8250\\
V396~HER   & 5982 -- 6002  & 5518 -- 5953               & 5518 -- 6002\\ 
          &               & 7347 -- 8057               & 7300 -- 7990 \\
Ton~256    & 5748 -- 5812  & 5408 -- 5564               & 5385 -- 5812\\   
          &               & 7218 -- 7582               & 7100 -- 7700\\      
\hline \\
\end{tabular}
\end{table}
The integration limits of the continuum and of individual broad emission lines
are also shown in the mean AGN spectra
(Figs.~\ref{fig:lk_1},\ref{fig:lk_2}).
The continuum region at 5100\,\AA\ is free of strong emission
and/or absorption lines. 
The intensities of the continuum flux at 5100\,\AA\
and of the integrated Balmer line intensities are given in Table~5.
The narrow line components were subtracted earlier. 
\begin{table}
\caption{Continuum fluxes at 5100\,\AA\ (in units of $10^{-15}$ \ergscma),
as well as integrated H$\beta$ and
H$\alpha$ line fluxes (in units $10^{-15}$ \ergscm) and the Balmer
decrement of our AGN sample. The continuum flux
at 6200\,\AA\ is also given for Mrk~926 (in parentheses).}
\tabcolsep+0.7mm
\begin{tabular}[H]{lcccc}
\hline \hline
\noalign{\smallskip}
Jul. Date & F$_{5100}$ & H$\beta$ &H$\alpha$\qquad{}&  H$\alpha$ / H$\beta$\\
2\,400\,000+\\
(1) & (2) & (3) & (4) & (5) \\
\noalign{\smallskip}
\hline
\noalign{\smallskip}
\multicolumn{2}{l}{3C~390.3} \\
\hline
      48089    & 2.24 $\pm{}$ 0.1   & 350  $\pm{}$ 28    & 1910 $\pm{}$ 230  & 5.46   $\pm{}$ 0.66\\
      48475    & 4.03 $\pm{}$ 0.2   & 496  $\pm{}$ 40    & 1830 $\pm{}$ 220  & 3.69   $\pm{}$ 0.44\\
      48866    & 2.23 $\pm{}$ 0.1   & 383  $\pm{}$ 31    & 1800 $\pm{}$ 220  & 4.70   $\pm{}$ 0.57\\
      49595    & 1.88 $\pm{}$ 0.1   & 279  $\pm{}$ 22    & 1310 $\pm{}$ 160  & 4.70   $\pm{}$ 0.57\\
\hline                         
\multicolumn{2}{l}{4U~0241+61}\\                   
\hline                         
      48475    & 0.79 $\pm{}$ 0.1   & 36.8 $\pm{}$ 4     & 546  $\pm{}$ 76  & 14.84  $\pm{}$ 2.07 \\
      48861    & 0.60 $\pm{}$ 0.1   & 29.7 $\pm{}$ 3     & 365  $\pm{}$ 51  & 12.29  $\pm{}$ 1.72 \\
      49596    & 0.71 $\pm{}$ 0.1   & 34.2 $\pm{}$ 3     & 476  $\pm{}$ 67  & 13.92  $\pm{}$ 1.96 \\
\hline                         
\multicolumn{2}{l}{E~1821+643}\\                   
\hline                         
      48095    & 6.67 $\pm{}$ 1.0   & 1150 $\pm{}$ 92    & 4630 $\pm{}$ 560  & 4.03   $\pm{}$ 0.49\\ 
      48476    & 7.34 $\pm{}$ 1.1   & 1050 $\pm{}$ 84    & 4430 $\pm{}$ 530  & 4.22   $\pm{}$ 0.50\\ 
      48813    & 6.87 $\pm{}$ 1.0   & 1130 $\pm{}$ 90    & 4310 $\pm{}$ 520  & 3.81   $\pm{}$ 0.46\\ 
      48860    & 6.87 $\pm{}$ 1.0   & 1150 $\pm{}$ 92    & 4860 $\pm{}$ 580  & 4.23   $\pm{}$ 0.50\\  
      49597    & 8.17 $\pm{}$ 1.2   & 1150 $\pm{}$ 92    & 4790 $\pm{}$ 570  & 4.17   $\pm{}$ 0.50\\  
\hline                         
 \multicolumn{2}{l}{Mrk~876} \\                    
\hline                         
      48093    & 3.32 $\pm{}$ 0.2   & 366  $\pm{}$ 29    & 1330 $\pm{}$ 160  & 3.63   $\pm{}$ 0.44\\  
      48474    & 3.55 $\pm{}$ 0.2   & 323  $\pm{}$ 26    & 1190 $\pm{}$ 140  & 3.68   $\pm{}$ 0.43\\  
      48814    & 4.06 $\pm{}$ 0.2   & 396  $\pm{}$ 32    & 1380 $\pm{}$ 170  & 3.48   $\pm{}$ 0.43\\  
      48862    & 4.28 $\pm{}$ 0.2   & 399  $\pm{}$ 32    & 1390 $\pm{}$ 170  & 3.48   $\pm{}$ 0.43\\  
      49598    & 4.15 $\pm{}$ 0.2   & 341  $\pm{}$ 27    & 1330 $\pm{}$ 160  & 3.90   $\pm{}$ 0.47\\  
\hline                         
\multicolumn{2}{l}{Mrk~926}\\                      
\hline                         
      48090    & 3.04 $\pm{}$ 0.3   & 434  $\pm{}$ 35    & 1400 $\pm{}$ 170 & 3.23   $\pm{}$ 0.39 \\                                          
               &(2.29 $\pm{}$ 0.2)  &              &             \\
      48092    &(2.03 $\pm{}$ 0.2)  &              & 1340 $\pm{}$ 160  \\                                          
      48478    & 4.11 $\pm{}$ 0.4   & 503  $\pm{}$ 40    &             \\                                  
      48480    &(2.8 $\pm{}$ 0.3)   &              & 1610 $\pm{}$ 190 & 3.20   $\pm{}$ 0.38 \\                                  
      48816    & 2.94 $\pm{}$ 0.3   & 469  $\pm{}$ 37    & 1920 $\pm{}$ 230 & 4.09   $\pm{}$ 0.49 \\                               
               &(2.55 $\pm{}$ 0.3 ) &              &             \\ 
      49215    & 2.03 $\pm{}$  0.2  & 292  $\pm{}$ 23    &             \\ 
      49577    &(1.35 $\pm{}$ 0.1)  &              & 761  $\pm{}$ 91   \\ 
      49596    & 2.1 $\pm{}$  0.2   & 198  $\pm{}$ 16    & 709  $\pm{}$ 85  & 3.58   $\pm{}$ 0.43 \\                                          
      50395    &(1.36 $\pm{}$ 0.1 ) &              & 685  $\pm{}$ 82   \\
      50635    &(1.5 $\pm{}$ 0.2)   &              & 689  $\pm{}$ 83   \\ 
\hline                         
\multicolumn{2}{l}{NGC~6814}\\                     
\hline                         
      48812    & 2.05 $\pm{}$ 0.1   & 55.9 $\pm{}$ 5     & 380  $\pm{}$ 46  & 6.80   $\pm{}$ 0.82 \\
      48866    & 3.45 $\pm{}$ 0.2   & 42.7 $\pm{}$ 3     & 360  $\pm{}$ 42  & 8.43   $\pm{}$ 0.98 \\
\hline                         
\multicolumn{2}{l}{OX~169} \\                      
\hline                         
      48093    & 1.10 $\pm{}$ 0.1   & 88.7 $\pm{}$ 4     & 374  $\pm{}$ 30  & 4.22   $\pm{}$ 0.34 \\
      48865    & 1.29 $\pm{}$ 0.1   & 96.2 $\pm{}$ 5     & 398  $\pm{}$ 30  & 4.14   $\pm{}$ 0.31 \\
      49598    & 1.31 $\pm{}$ 0.1   & 94.1 $\pm{}$ 5     & 396  $\pm{}$ 30  & 4.21   $\pm{}$ 0.32 \\
\hline                         
\multicolumn{2}{l}{PKS~2349-014}\\                
\hline                         
      48094    & 1.42 $\pm{}$ 0.1   & 147  $\pm{}$ 12    & 512   $\pm{}$ 15 & 3.48   $\pm{}$ 0.10 \\
      48864    & 0.68 $\pm{}$ 0.03  & 74.3 $\pm{}$ 6     & 289   $\pm{}$ 10 & 3.89   $\pm{}$ 0.13 \\
      49598    & 1.59 $\pm{}$ 0.1   & 120  $\pm{}$ 10    & 306   $\pm{}$ 37 & 2.55   $\pm{}$ 0.31 \\    
\hline                         
\multicolumn{2}{l}{V396~Her}  \\                   
\hline                         
      48093    & 0.79 $\pm{}$ 0.04  & 45.9 $\pm{}$ 4     & 203  $\pm{}$ 24  & 4.72   $\pm{}$ 0.56 \\
      48864    & 0.35 $\pm{}$ 0.02  & 24.1 $\pm{}$ 2     & 101  $\pm{}$ 12  & 4.58   $\pm{}$ 0.54 \\
      49596    & 0.47 $\pm{}$ 0.02  & 31.3 $\pm{}$ 3     & 120  $\pm{}$ 15  & 4.42   $\pm{}$ 0.52 \\
\hline
\multicolumn{2}{l}{TON~256} \\                     
\hline                         
      48094    & 1.00 $\pm{}$ 0.1   & 144   $\pm{}$ 12   & 680  $\pm{}$ 80  & 4.19   $\pm{}$ 0.50 \\
      48475    & 1.08 $\pm{}$ 0.1   & 149   $\pm{}$ 12   & 683  $\pm{}$ 80  & 3.83   $\pm{}$ 0.48 \\
\noalign{\smallskip}
\hline
\noalign{\smallskip}
\end{tabular}
\end{table}
\subsection{Light curves and mean spectra}
\begin{figure*}
  \hbox{\includegraphics[width=80mm, height=48mm]{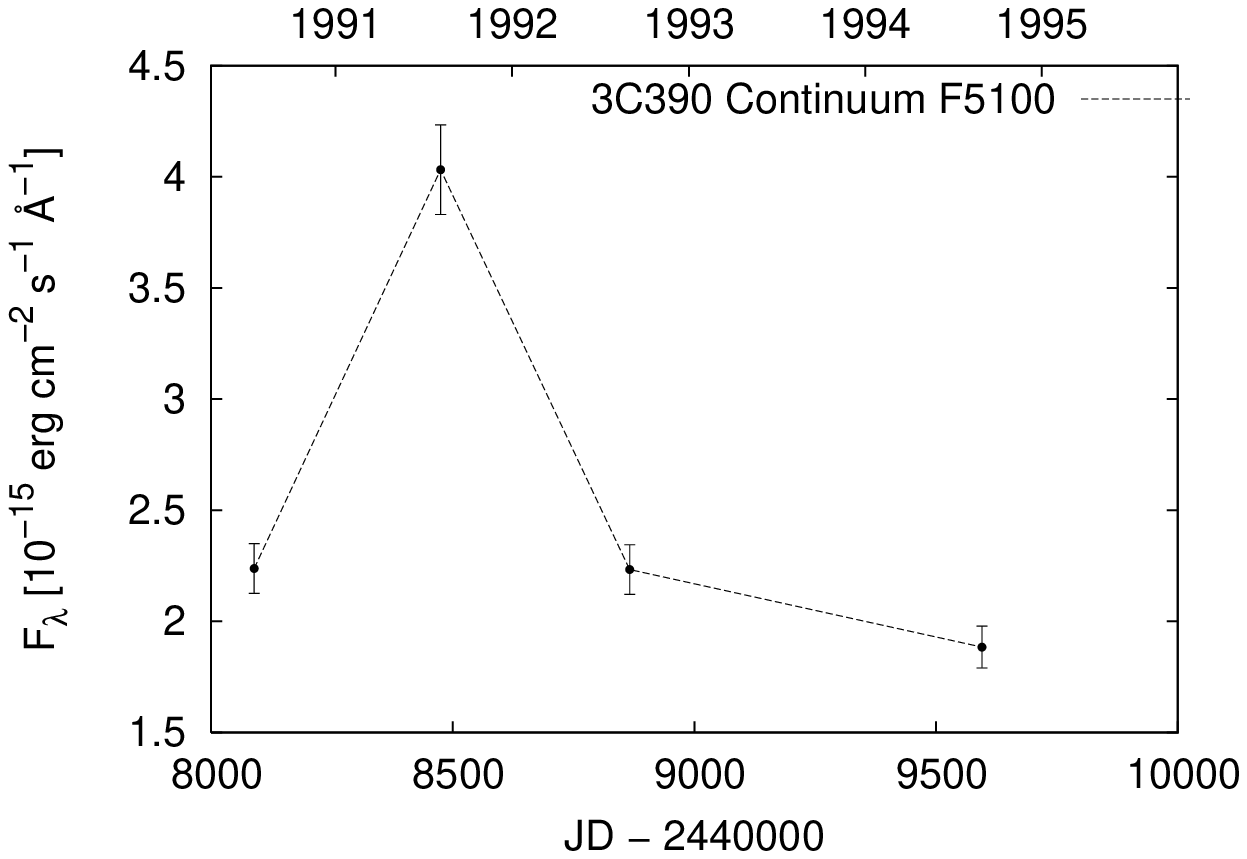}\hspace*{-2mm}
        \includegraphics[width=80mm, height=48mm]{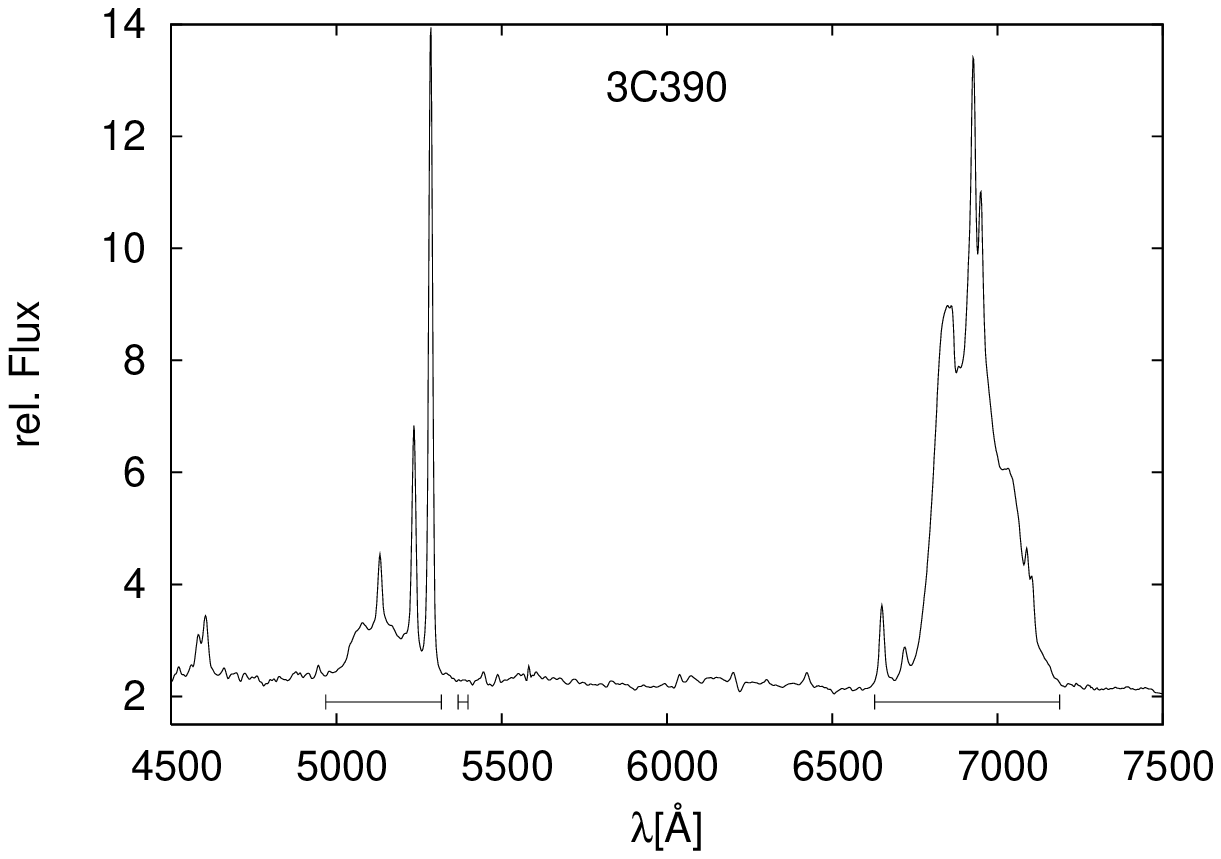}}
  \hbox{\includegraphics[width=80mm, height=48mm]{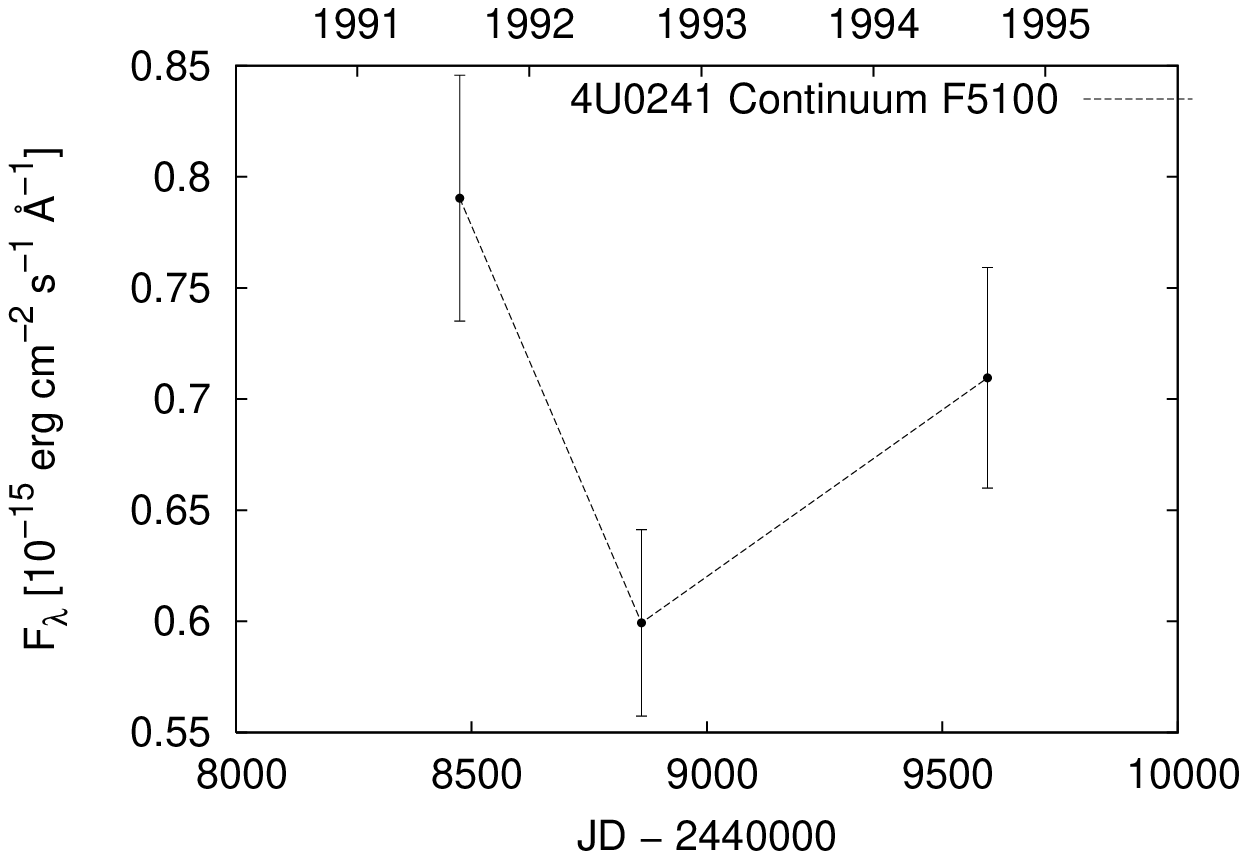}\hspace*{-2mm}
        \includegraphics[width=80mm, height=48mm]{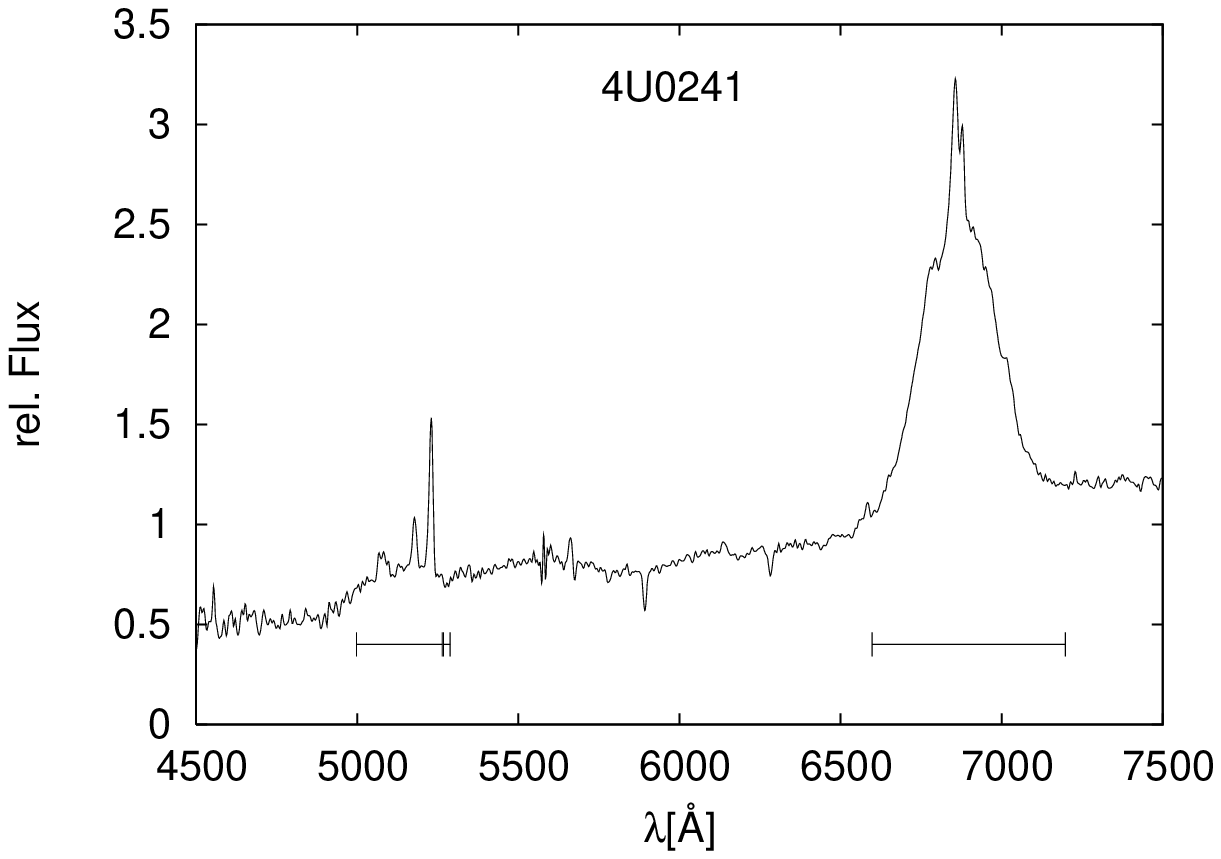}}
  \hbox{\includegraphics[width=80mm, height=48mm]{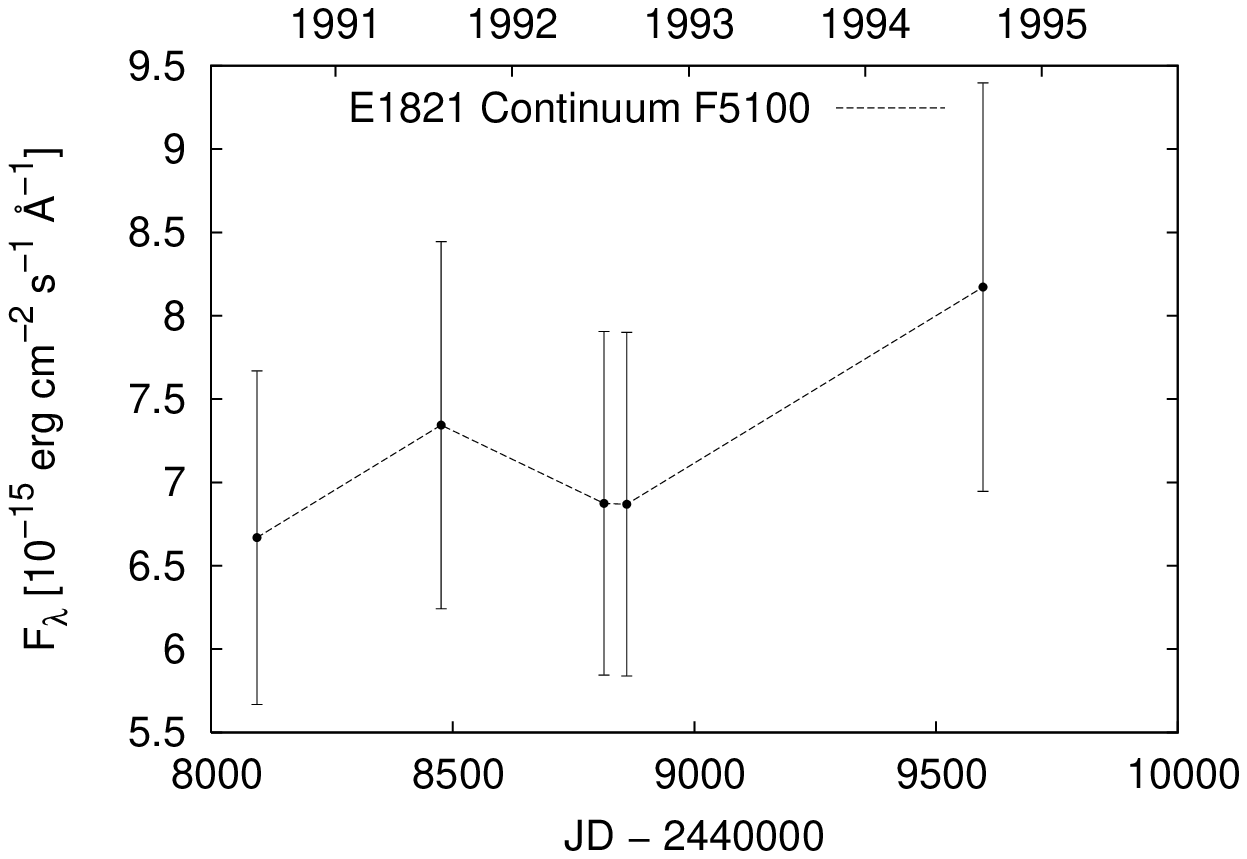}\hspace*{-2mm}
        \includegraphics[width=80mm, height=48mm]{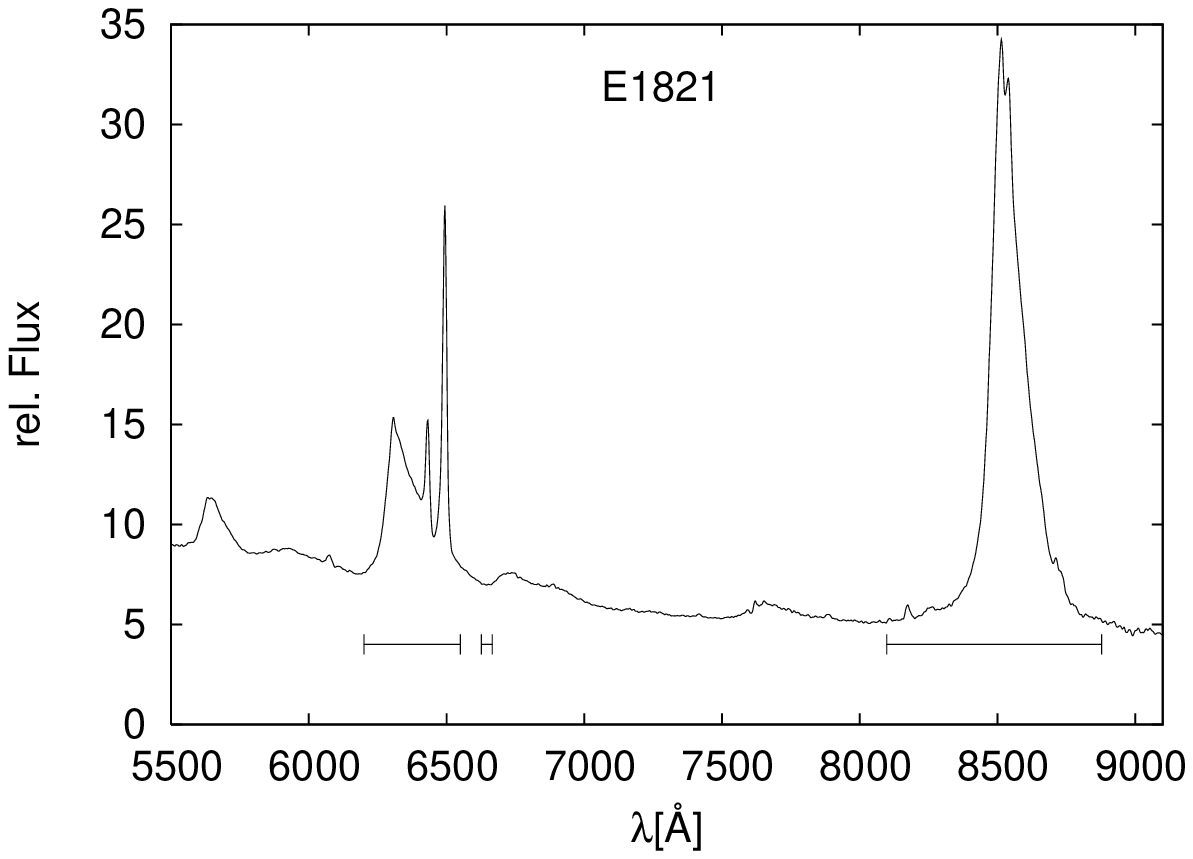}}
  \hbox{\includegraphics[width=80mm, height=48mm]{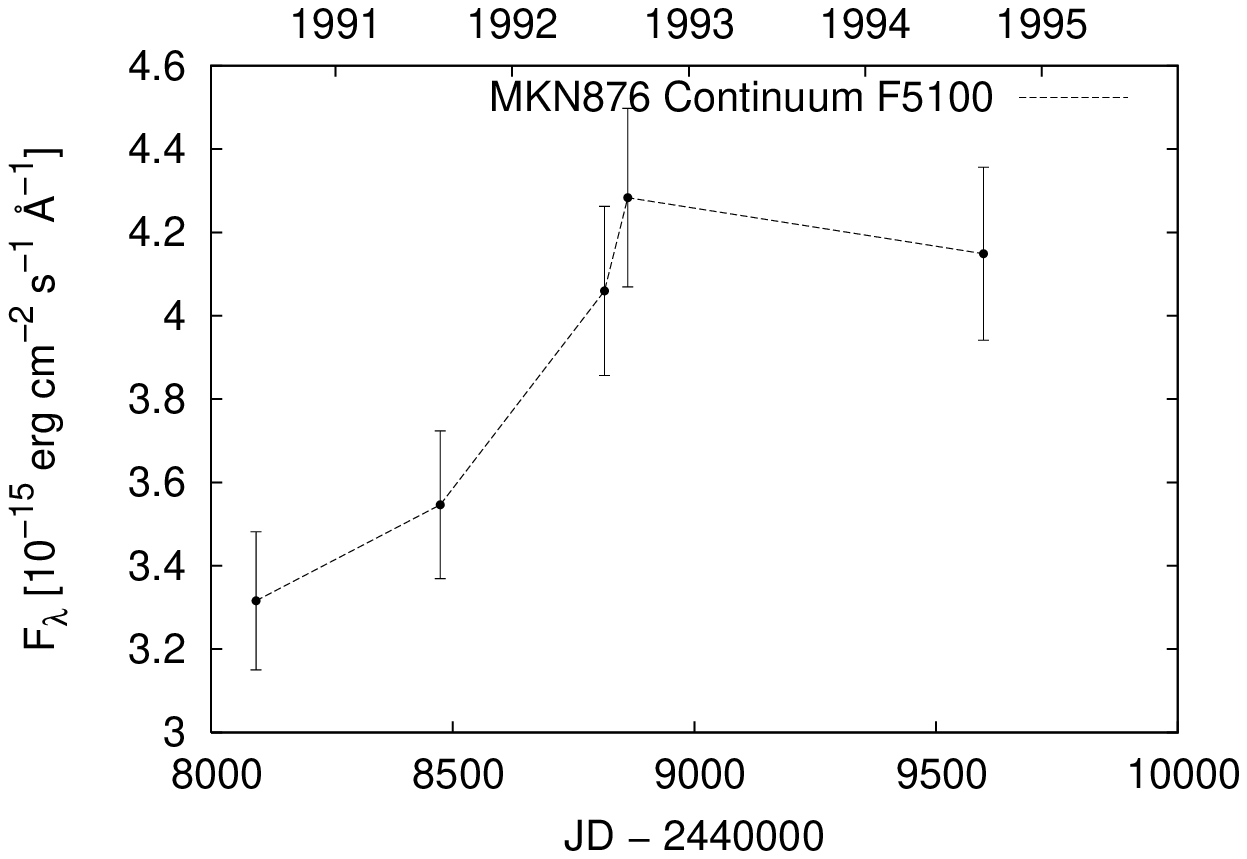}\hspace*{-2mm}
        \includegraphics[width=80mm, height=48mm]{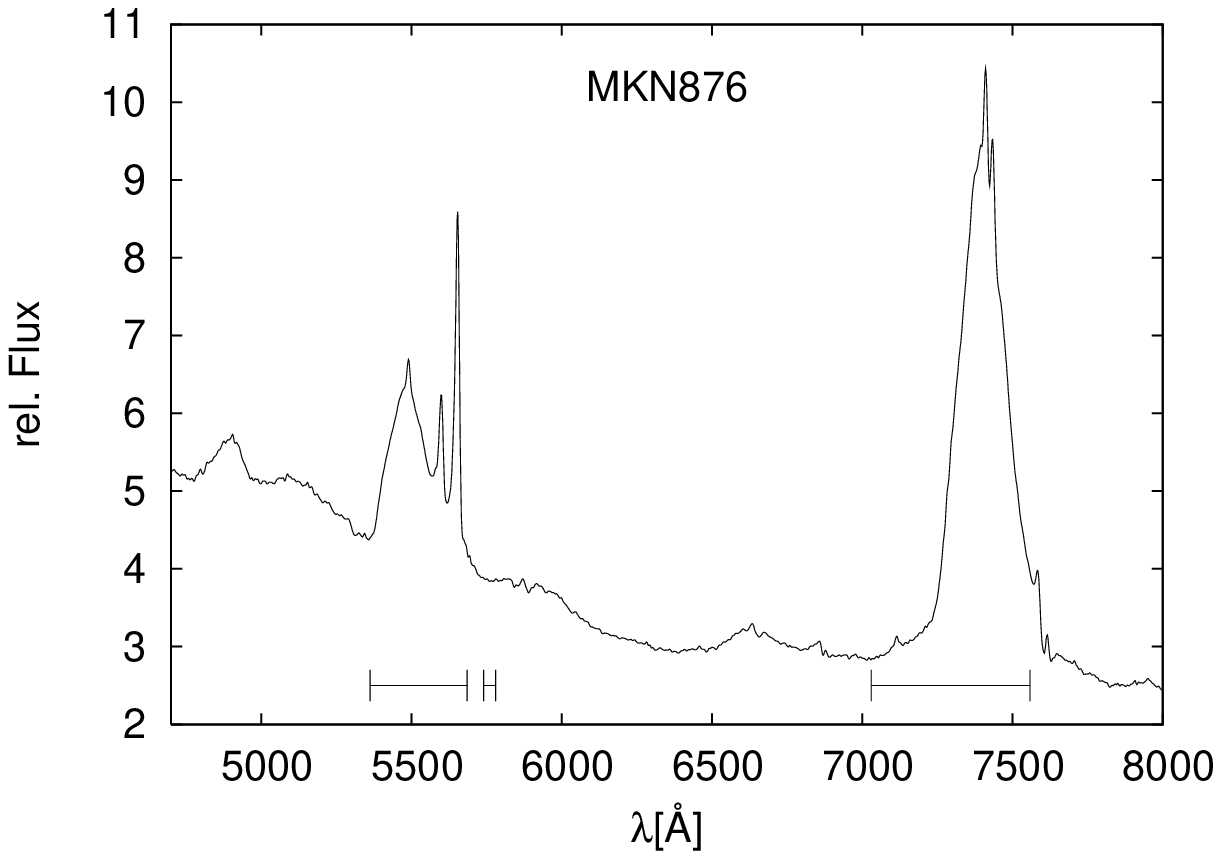}}
  \hbox{\includegraphics[width=80mm, height=48mm]{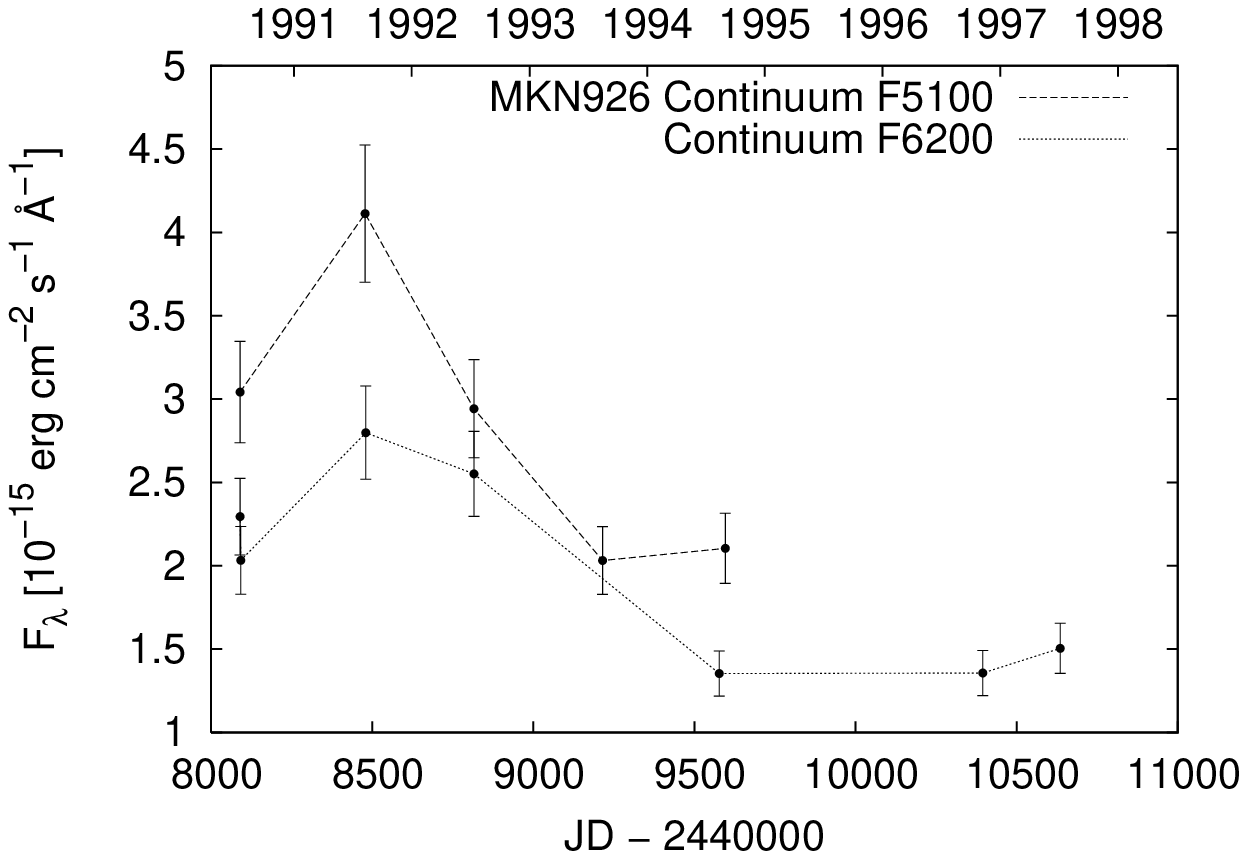}\hspace*{-2mm}
        \includegraphics[width=80mm, height=48mm]{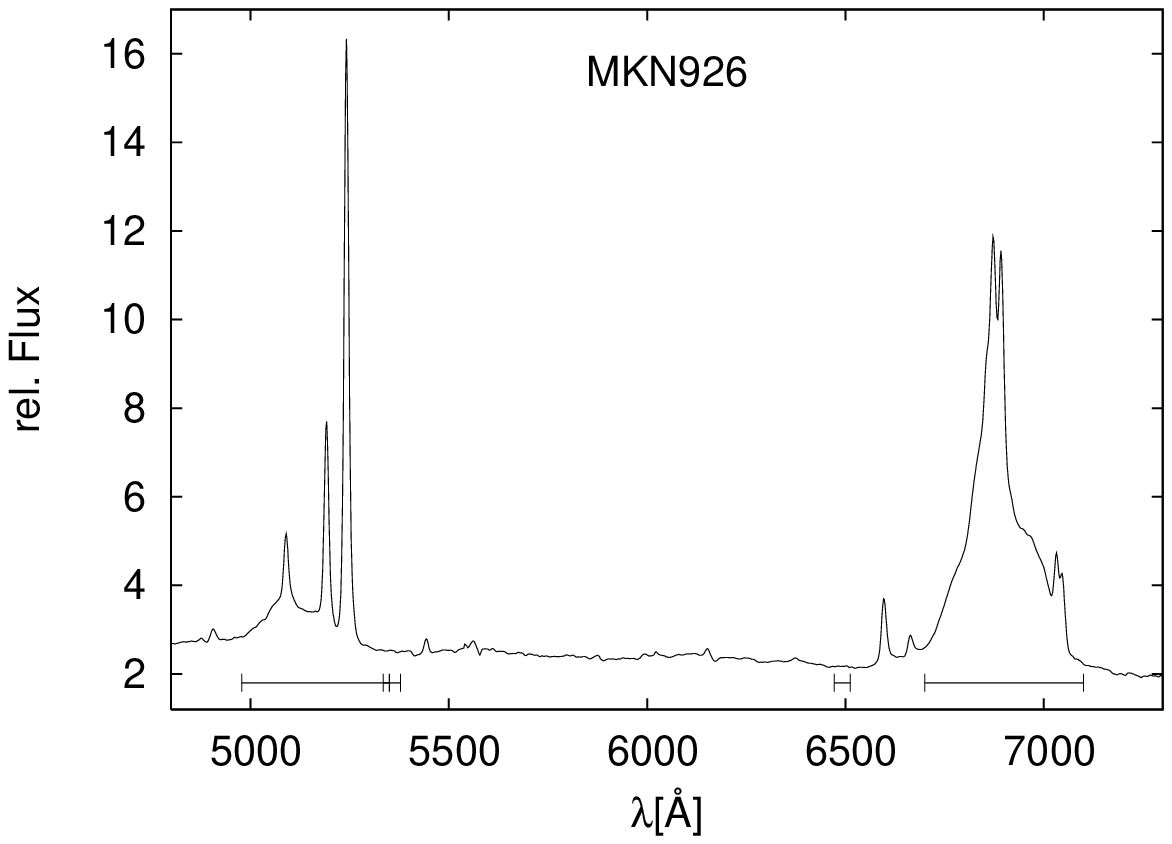}}
        \vspace*{5mm} 
   \caption{Light curves of the continuum flux at 5100\,\AA\ (and at 6200\,\AA\
for Mrk~926)
  (in units of 10$^{-15}$ erg cm$^{-2}$ s$^{-1}$\,\AA$^{-1}$) and
mean AGN spectra.
   The points in the light curves
   are connected by a dotted line to aid the eye.}
  \label{fig:lk_1}
\end{figure*}

\begin{figure*}
  \hbox{\includegraphics[width=80mm, height=48mm]{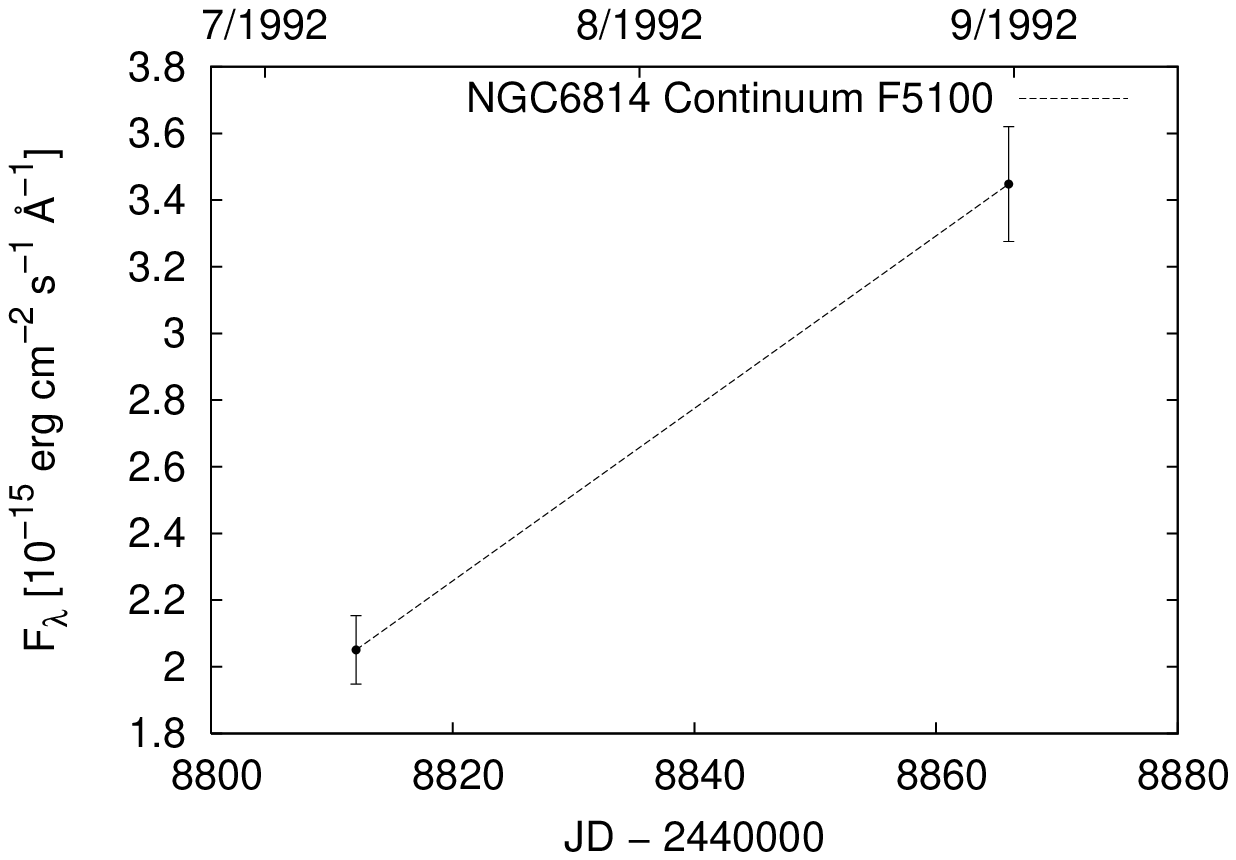}\hspace*{-2mm}
        \includegraphics[width=80mm, height=48mm]{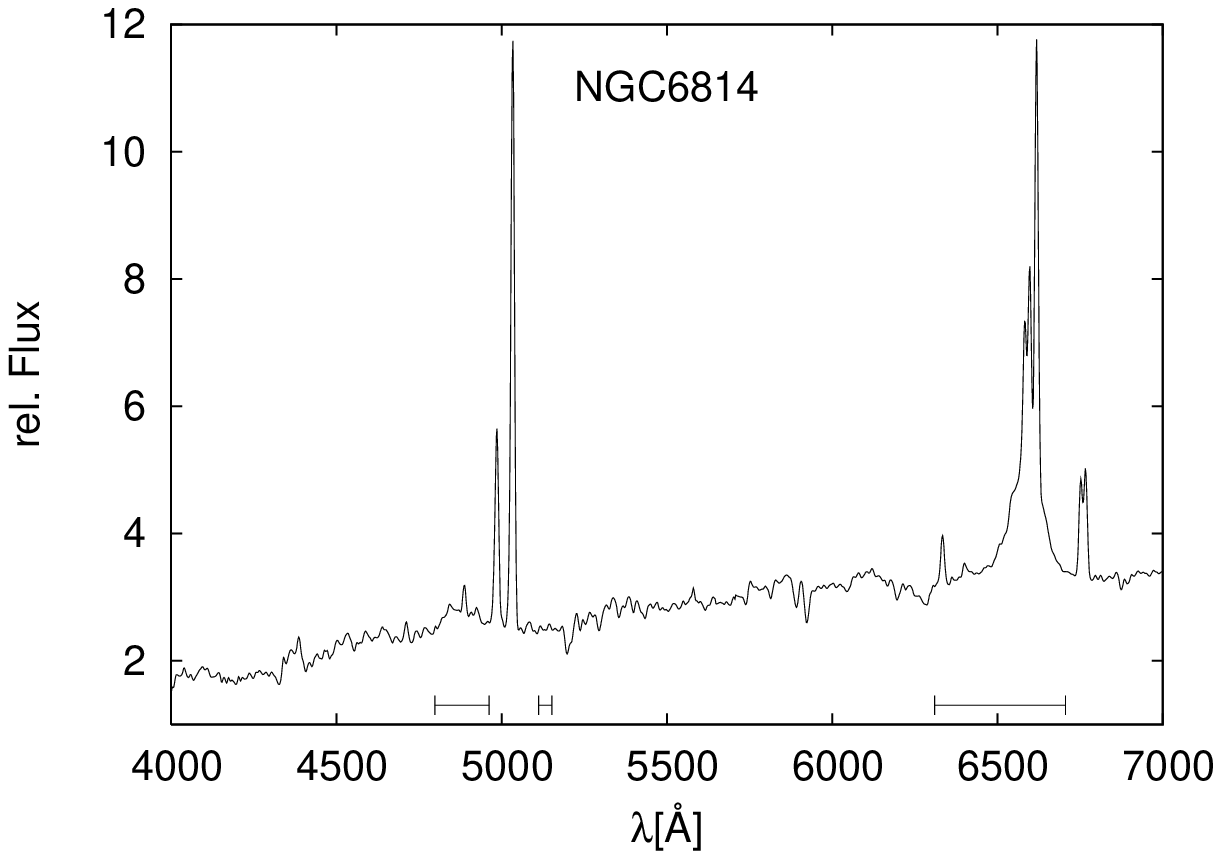}}
  \hbox{\includegraphics[width=80mm, height=48mm]{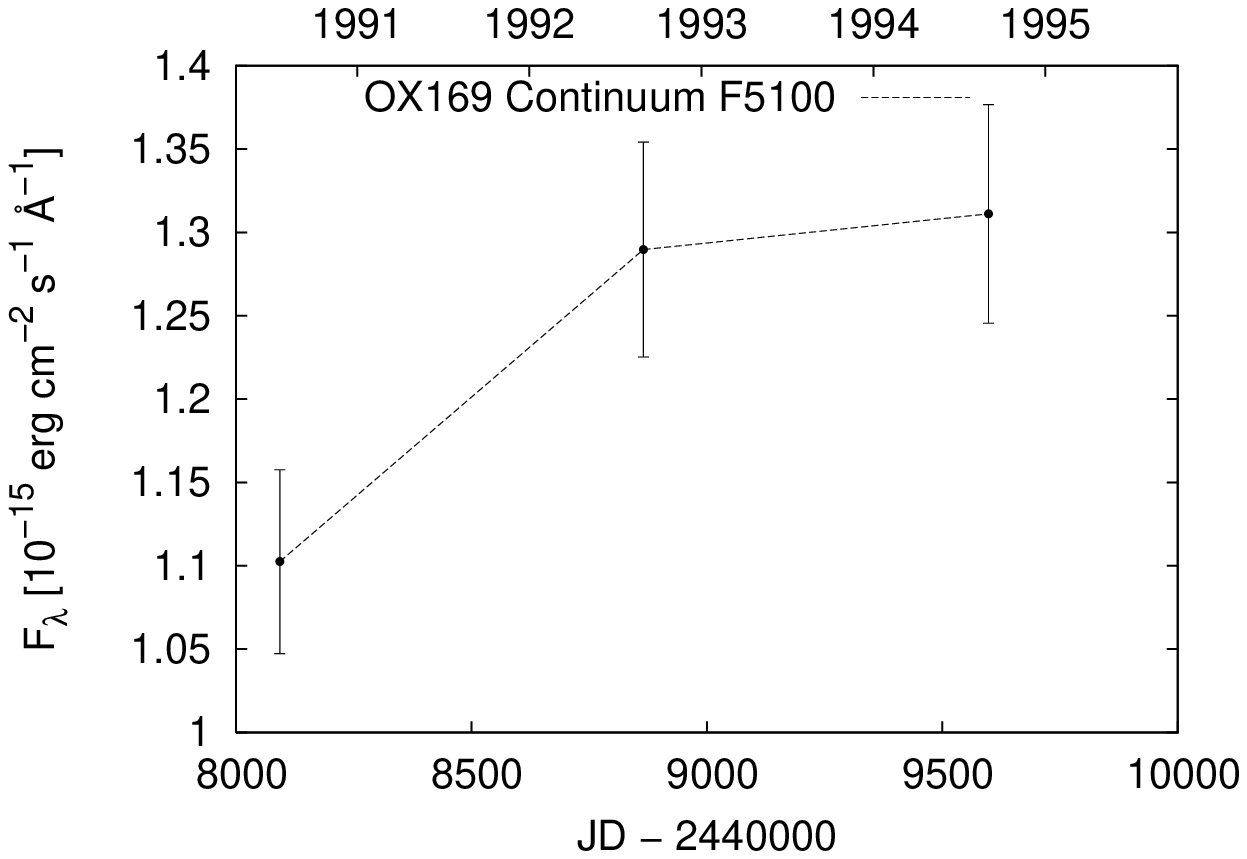}\hspace*{-2mm}
        \includegraphics[width=80mm, height=48mm]{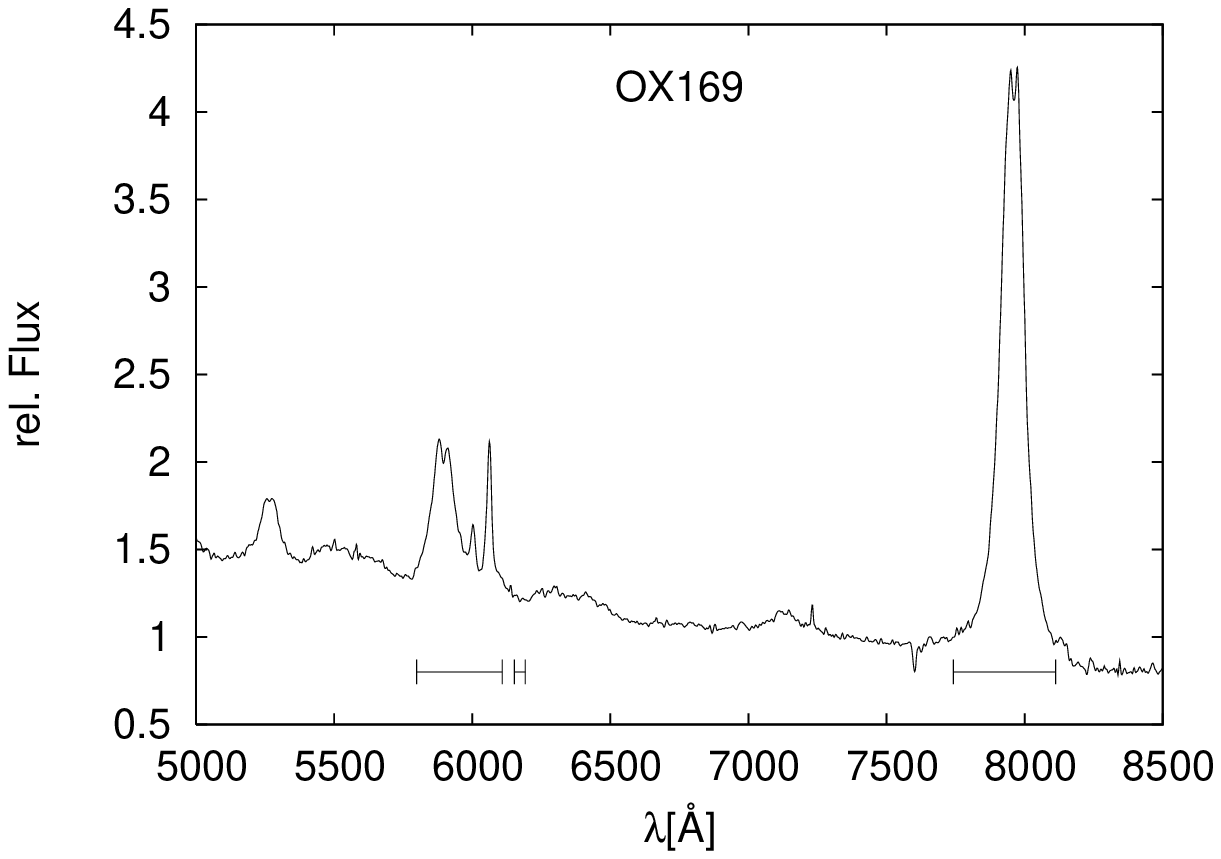}}
  \hbox{\includegraphics[width=80mm, height=48mm]{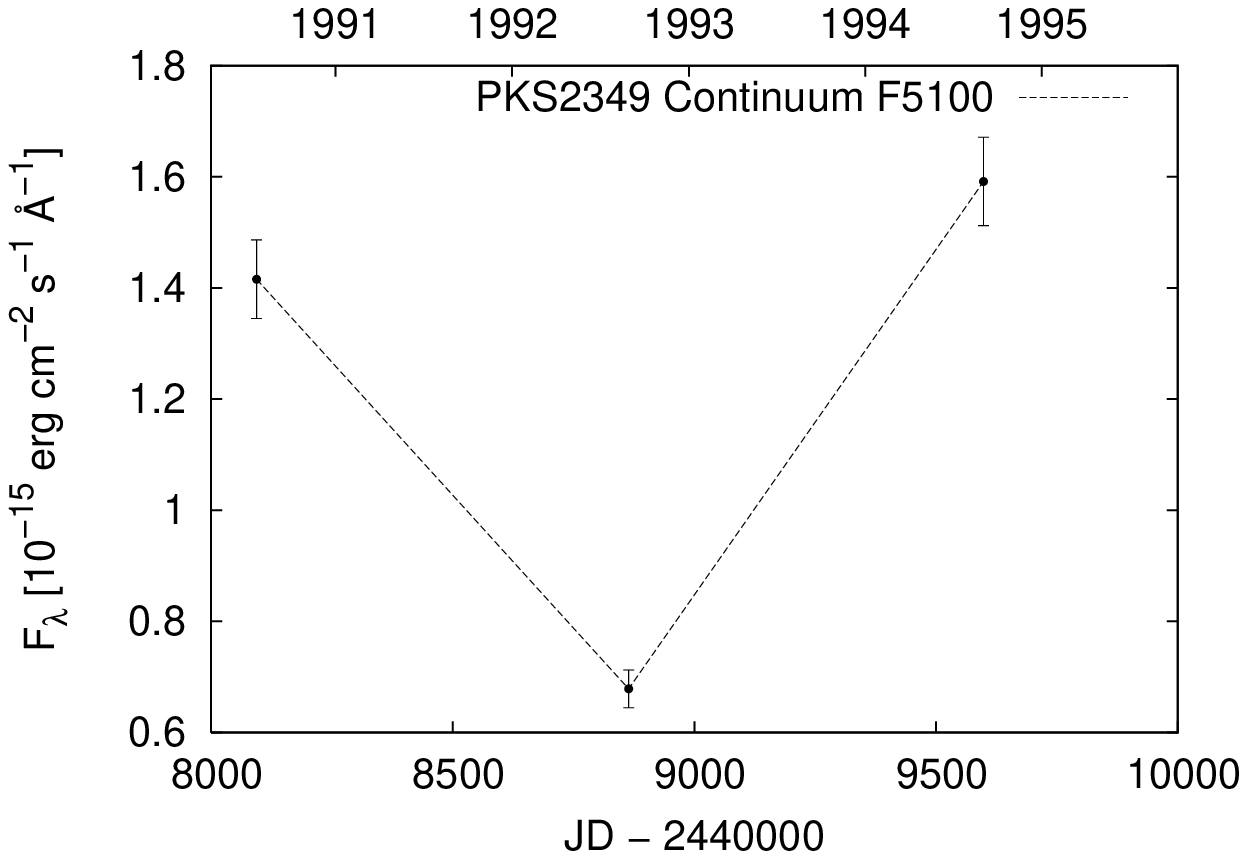}\hspace*{-2mm}
        \includegraphics[width=80mm, height=48mm]{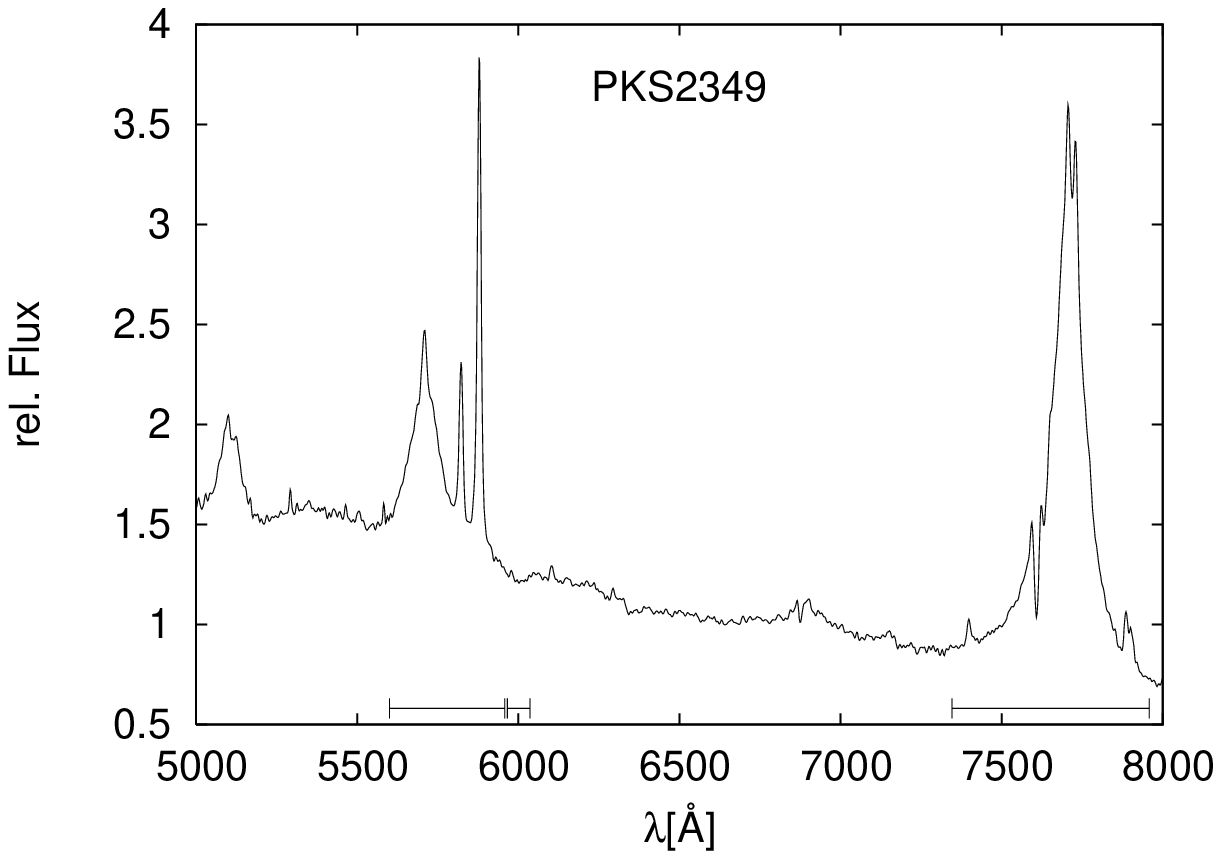}}
  \hbox{\includegraphics[width=80mm, height=48mm]{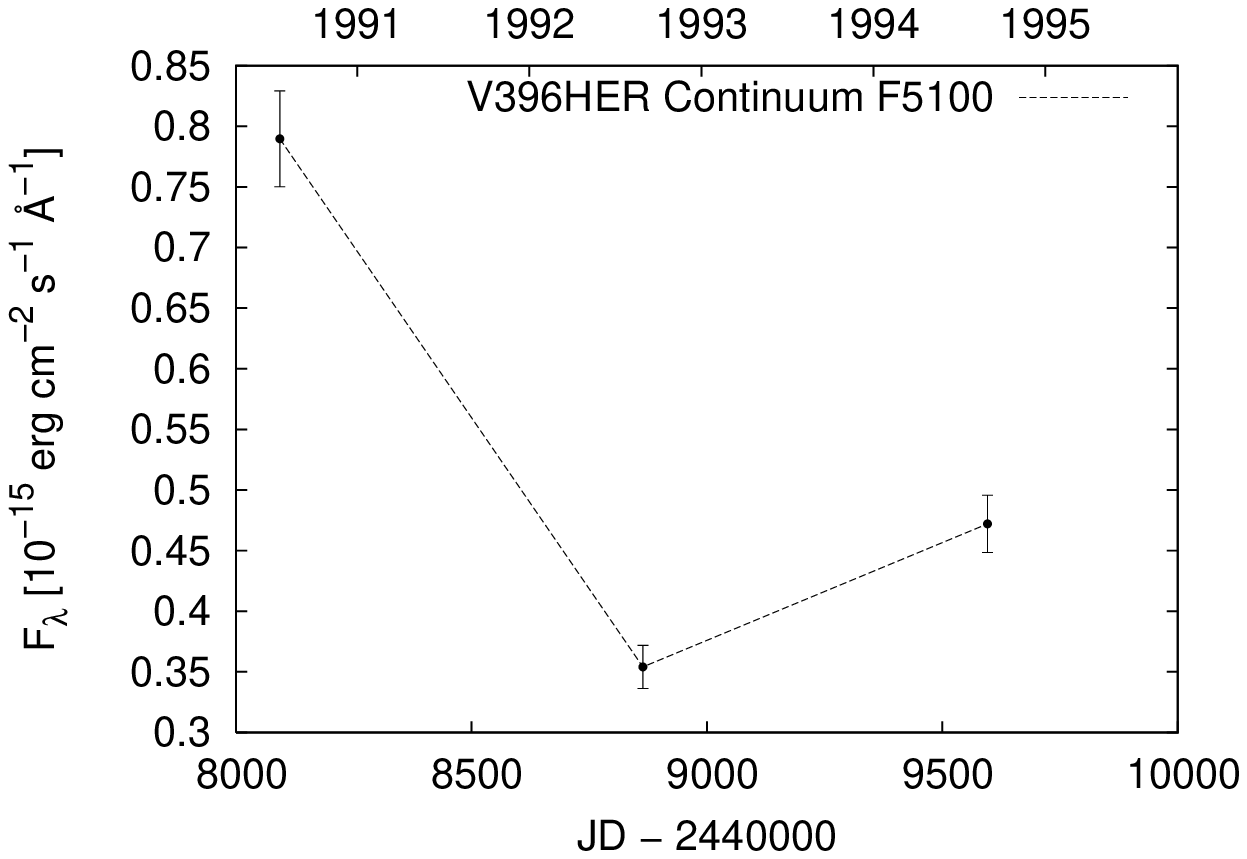}\hspace*{-2mm}
        \includegraphics[width=80mm, height=48mm]{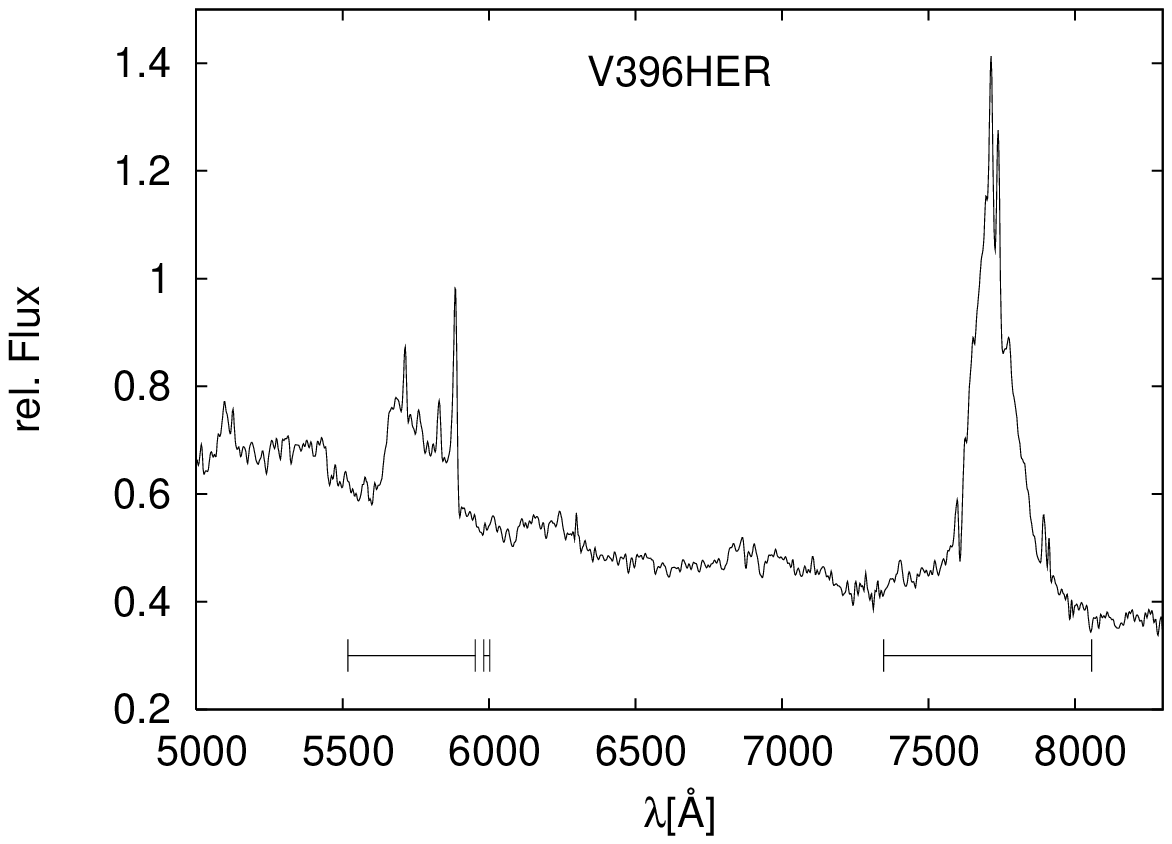}}
  \hbox{\includegraphics[width=80mm, height=48mm]{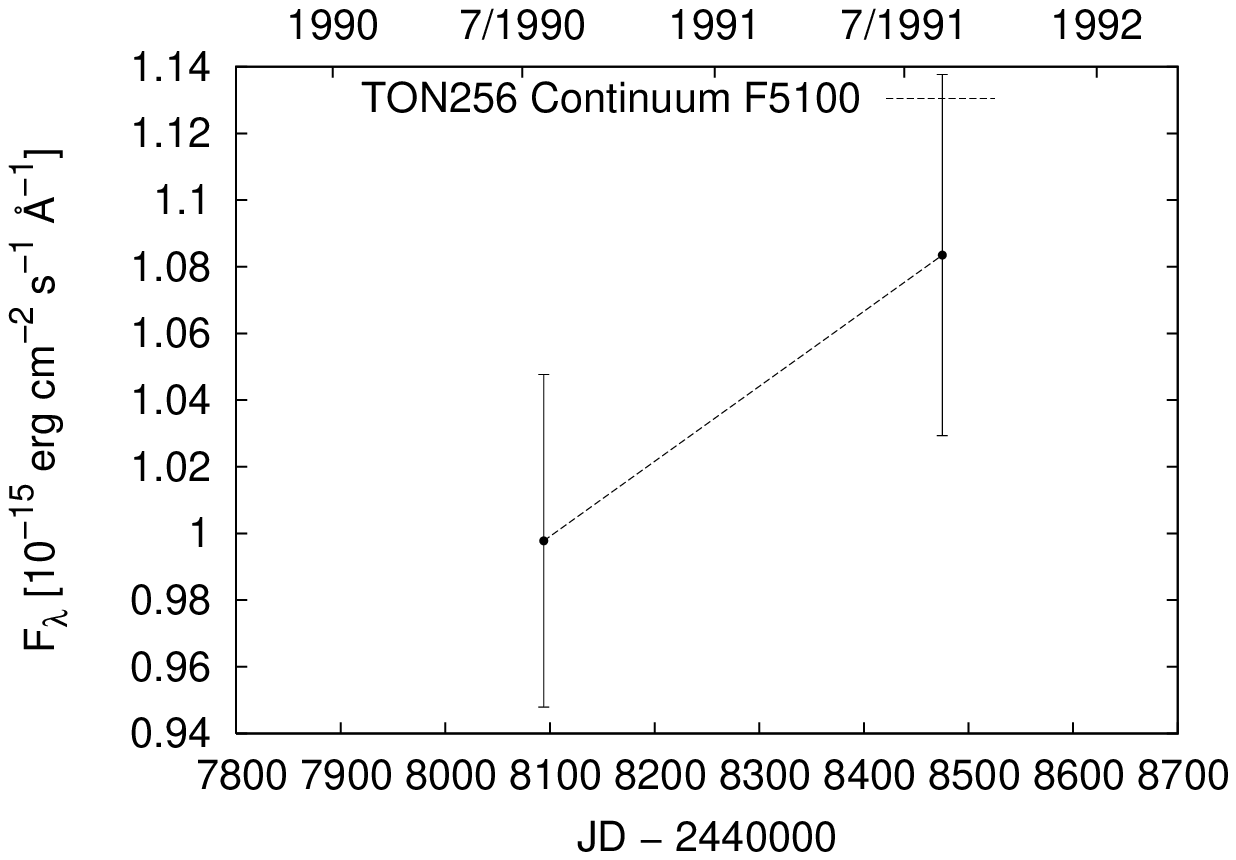}\hspace*{-2mm}
        \includegraphics[width=80mm, height=48mm]{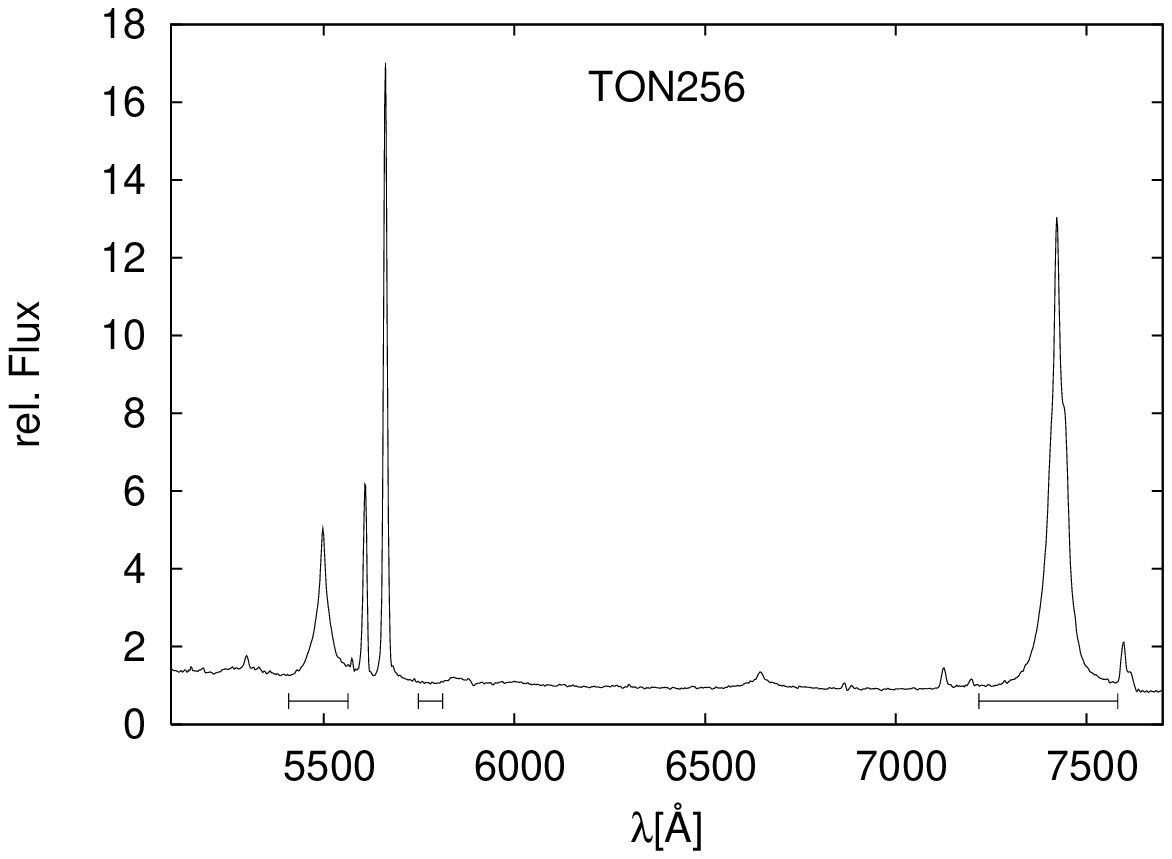}}
        \vspace*{5mm} 
  \caption{Same as Fig.~\ref{fig:lk_1}}
  \label{fig:lk_2}
\end{figure*}

The continuum light curves of our sample of
very broad-line AGNs
are shown in Figs.~\ref{fig:lk_1},\ref{fig:lk_2}.
The Julian Date (bottom), as well as year and month (top),
of observations are given with the plots.
The continuum flux was measured at 5100~\AA (rest-frame).
The light curve of the continuum flux at 6200\,\AA\
is shown for Mrk~926 in addition, 
because at some epochs we only took spectra of the H$\alpha$ region
for this galaxy.
The calculated mean spectra of the
very broad-line AGNs are also presented in Figs.~\ref{fig:lk_1},\ref{fig:lk_2}.
We determined
the line-widths (FWHM) of the broad emission lines and their
equivalent values in addition to the continuum intensities.
These results are given in Table~6.
\subsection{Variability analysis}
We determined the amplitudes
of the continuum flux variations
and of the broad emission-line intensities as the next steps.
We concentrated on the fractional variations.
A definition of the fractional variation $F_{var}$ is given by
Rodr{\'\i}guez-Pascual et al.\ (\cite{rodriguez97}):
\begin{equation}
  \label{eq:rod}
F_{var} = \frac{\sqrt{{\sigma_F}^2 - \Delta^2}}{<F>}
\end{equation}
where $<$F$>$ is the mean flux
over the period of observations, $\sigma_F$ the
standard deviation, and $\Delta^2$ the mean square value of the
uncertainties $\Delta_{i}$ associated with the fluxes $f_{i}$. 
We list the statistics of the continuum and
H$\beta$ line variations that we derived for our objects in Table~7.
\begin{table*}
\caption{Continuum and H$\beta$ line
variability statistics, mean H$\beta$ line widths (FWHM),
and H$\beta$ equivalent widths, as well as minimum and maximum
observed values for the continuum and H$\beta$
of our AGN sample.}
\tabcolsep+2.2mm
\begin{tabular}{lcrcccccc}
\hline
\noalign{\smallskip}
Object        &   F$_{var}$ F$_{5100}$ &  F$_{var}$ H$\beta$ & FWHM  H$\beta$ & EQW  H$\beta$ 
&  $F_{min}$ F$_{5100}$ & $F_{max}$ F$_{5100}$  & $F_{min}$ H$\beta$ & $F_{max}$ H$\beta$ \\
    &     &     &  [kms$^{-1}$] & [\AA{}] \\
(1) & (2) & (3) & (4) & (5)  & (6) & (7) & (8) & (9) \\
\hline
3C390         &   0.363      &  0.224      &  11531   $\pm{}$ 1440        & 111     $\pm{}$ 20          & 1.88 & 4.06 & 278.5 & 512.2  \\    
4U0241+61     &   0.118      &  0.037      &  8722    $\pm{}$ 930         & 43      $\pm{}$ 1           & 0.60 & 0.79 & 29.7  & 36.8   \\    
E1821+643     &   0.128      &  $<$ 0.020           &  6228    $\pm{}$ 351         & 116  $\pm{}$ 12    & 6.67 & 8.17 & 996.3 & 1150.8 \\     
Mrk876        &   0.086      &  0.043      &  7385    $\pm{}$ 448         & 79      $\pm{}$ 9           & 3.32 & 4.28 & 323.0 & 398.7  \\    
Mrk926        &   0.279      &  0.330      &  8505    $\pm{}$ 150         & 143     $\pm{}$ 26          & 2.03 & 4.11 & 198.1 & 502.7   \\   
NGC6814       &   0.319      &  0.124      &  7557    $\pm{}$ 353         & 25      $\pm{}$ 13          & 1.95 & 3.45 & 42.7  & 56.5   \\    
OX169         &   0.078      &  $<$ 0.020           &  5008    $\pm{}$ 294         & 59      $\pm{}$ 4  & 1.10 & 1.31 & 88.7  & 96.2   \\        
PKS2349-14    &   0.391      &  0.311      &  5132    $\pm{}$ 1082        & 83      $\pm{}$ 4           & 0.68 & 1.42 & 74.3  & 146.6   \\   
V396HER       &   0.415      &  0.319      &  12334   $\pm{}$ 262         & 48      $\pm{}$ 5           & 0.35 & 0.79 & 24.1 &  45.9   \\    
Ton256        &   0.030      &  $<$ 0.020           &  1458    $\pm{}$ 18          & 108     $\pm{}$ 4  & 1.0  & 1.08 & 143.7 & 149.3   \\       

\noalign{\smallskip}                                                            
\hline                                                                          
\noalign{\smallskip}
\end{tabular}
\end{table*}

For three of our galaxies (E1821+643, OX169, Ton256), we only get upper limits
of the H$\beta$ line variations.
In these cases
the H$\beta$ variations ${\sigma_F}^2$ were smaller than the error   
${\Delta}^2$ in Eq.~2. 
In this paper we focus only on the fractional variation of the
continuum and of the H$\beta$ line.

In a few cases, the variability behavior of more than one emission line
has been investigated: NGC5548, 3C390.3
 (O'Brien et al. \cite{OBrien98};  Dietrich et al.  \cite{dietrich98}; 
  Peterson \&  Wandel \cite{peterson99}),
  Mrk110 (Kollatschny et al.\cite{kollatschny01}). 
When considering different emission lines in one AGN,
the following trend was seen.
The broader (FWHM) lines show stronger
variability amplitudes, and the broader lines originate closer to the central
black hole.
\subsection{Correlation analysis and results}
\begin{figure}[hbp]
        \centering
        \includegraphics[width=8.5cm]{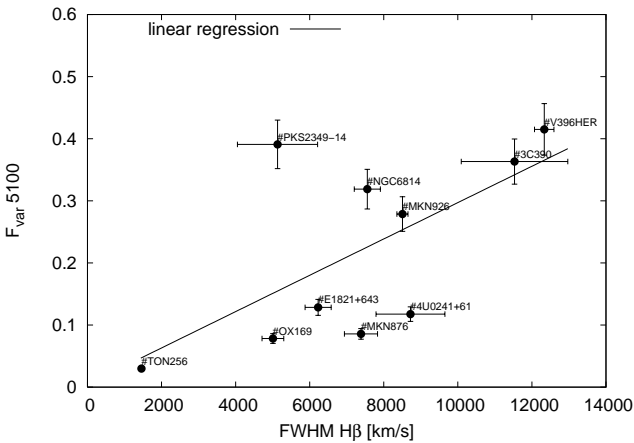}
        \caption{Fractional variation of the continuum at 5100\,\AA\
 versus H$\beta$ FWHM (very broad-line AGN sample).}
        \label{fig:fvar_5100_vs_hwb_hb_o2}
\end{figure}

\begin{figure}[htbp]
        \centering
        \includegraphics[width=8.5cm]{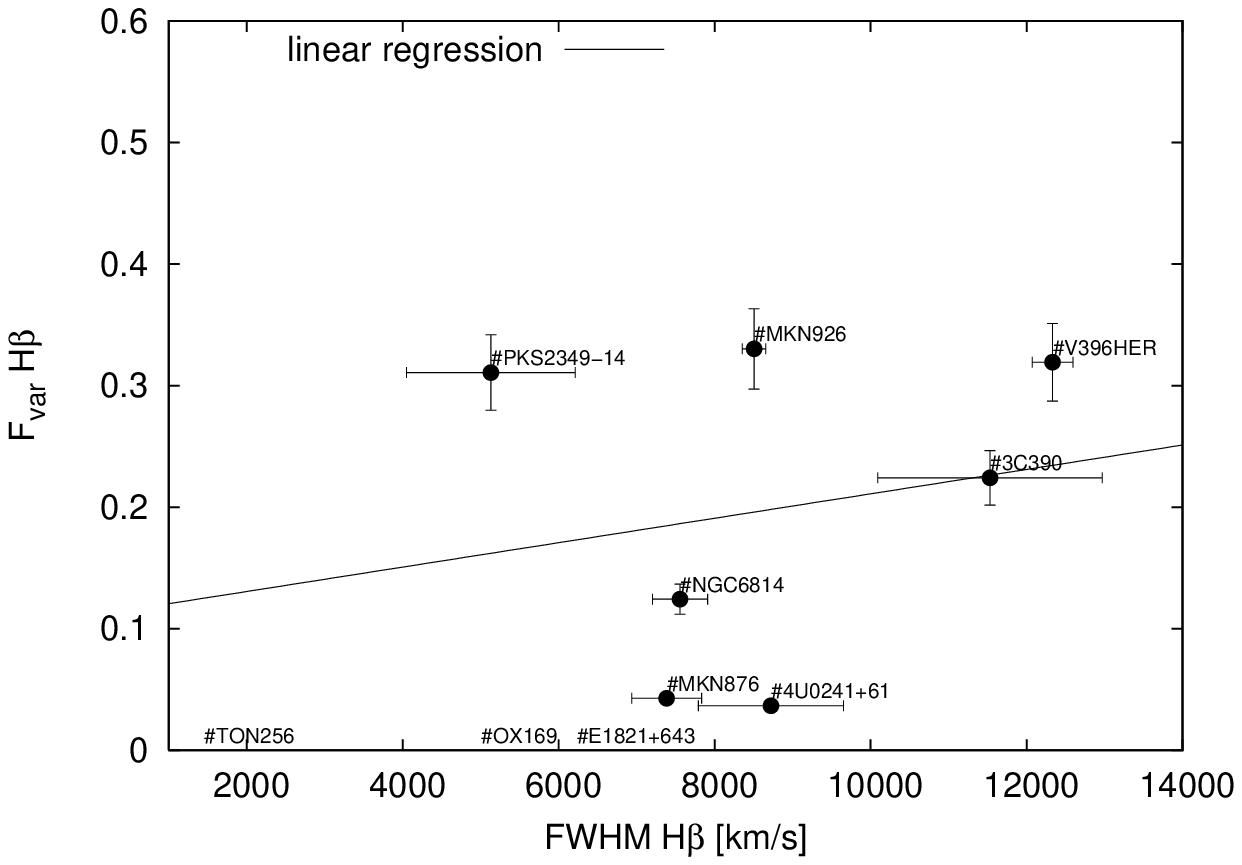}
        \caption{Fractional variation of H$\beta$ versus
          H$\beta$ FWHM (very broad-line AGN sample).}
        \label{fig:fvar_hb_vs_hwb_hb_o2}
\end{figure}

\begin{figure}[htbp]
        \centering
        \includegraphics[width=8.5cm]{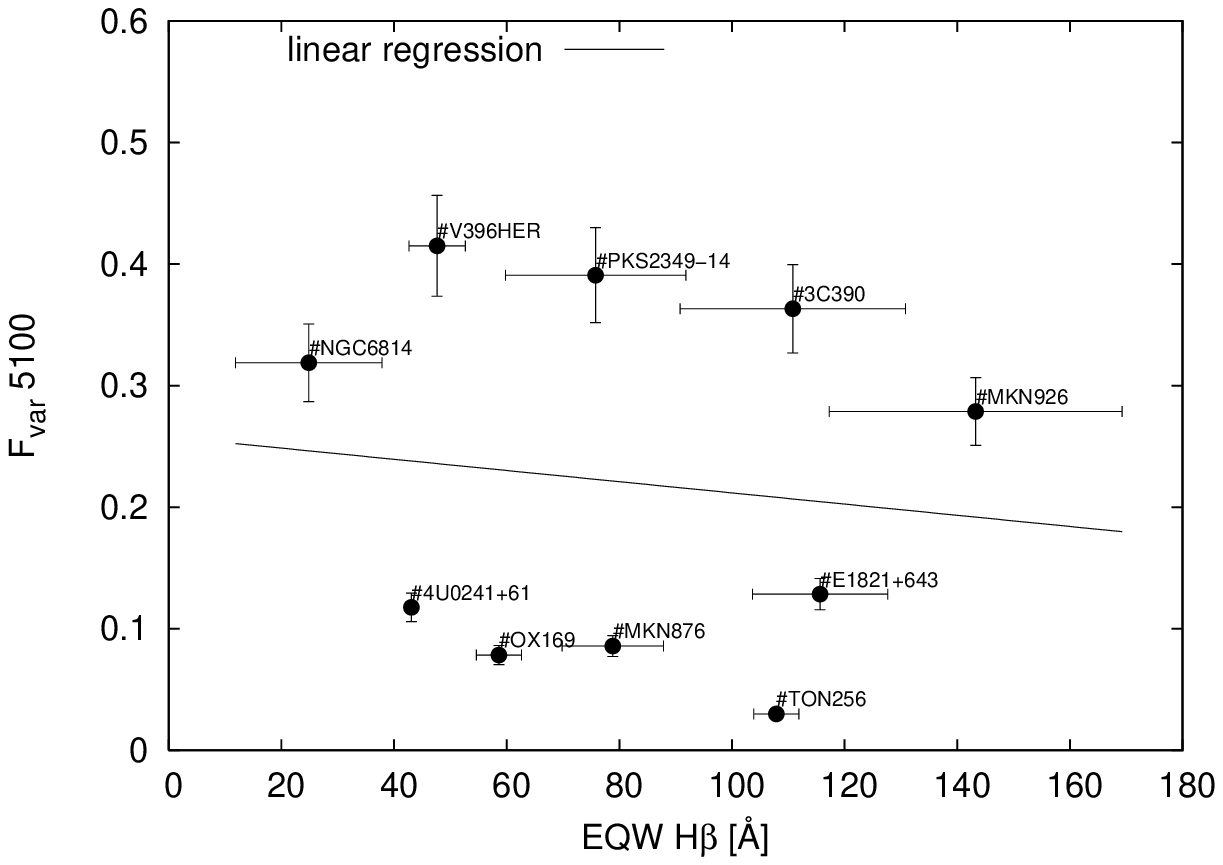}
        \caption{Fractional variation of the continuum at 5100\,\AA\ versus
          H$\beta$ equivalent width (very broad-line AGN sample).}
        \label{fig:fvar_5100_vs_eqw_hb_o2}
\end{figure}

\begin{figure}[htbp]
        \centering
        \includegraphics[width=8.5cm]{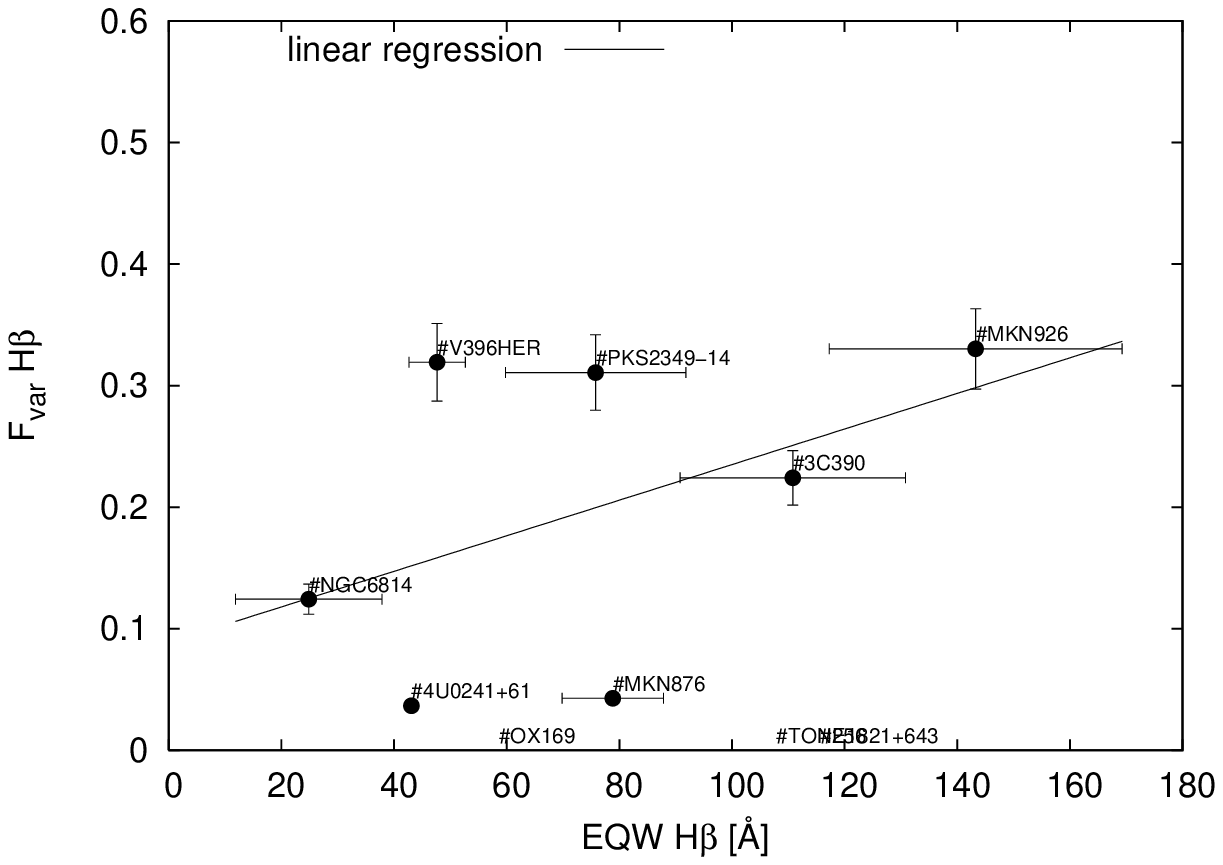}
        \caption{Fractional variation of H$\beta$ versus
          H$\beta$ equivalent width (very broad-line AGN sample).}
        \label{fig:fvar_hb_vs_eqw_hb_o2}
\end{figure}

In the current study of broad-line AGNs, we are looking for correlations 
between the variability amplitudes of the optical continuum and the H$\beta$ 
emission-line flux, $F_{var}(5100)$ and $F_{var}$(H$\beta$), with the 
emission-line profile width and the equivalent width of H$\beta$. 
The optical continuum and  H$\beta$ intensity
variations are plotted over 
the H$\beta$ line widths and their equivalent widths in
Figs. \ref{fig:fvar_5100_vs_hwb_hb_o2}--\ref{fig:fvar_hb_vs_eqw_hb_o2}.
The linear fit  we derived from the Pearson correlation  is overlaid 
in each figure. 

Besides the Pearson correlation coefficient $r_{p}$
\begin{equation}
  \label{eq:pearson}
  r_{p} = \frac{\sum\limits_{i=1}^{n}(x_{i}-\langle x\rangle)(y_{i}-\langle y\rangle)}
  {\sqrt{\sum\limits_{i=1}^{n}(x_{i}-\langle x\rangle)^{2} 
         \sum\limits_{i=1}^{n}(y_{i}-\langle y\rangle)^{2}}}
\end{equation}
we calculated for each relation
the Spearman's rank-correlation coefficient $r_{s}$
\begin{equation}
  \label{eq:spearman}
r_{s} = 1 - \frac{6\sum\limits_{i=1}^{n} (Rg(x_{i}) - Rg(y_{i}))^{2}}{n^{3}-n}~,
\end{equation}
as well as the Kendall correlation coefficient $r_{k}$
\begin{equation}
  \label{eq:kendall}
r_{k} = \frac{1}{n^{3}} \sum\limits_{i,j=1}^{n} sign(x_i-x_j) sign(y_i-y_j)~. 
\end{equation}
Here $P_{p}$, $P_{s}$, and $P_{k}$ are the associated probabilities for random
correlations in percent (Bevington, \cite{Bevington92},  Press et al.
\cite{press92}).
The Pearson correlation coefficient tests only a linear relation, while
the Spearman rank-correlation coefficient and 
Kendall correlation coefficient test for a general monotonic relation
(e.g. Press et al. \cite{press92}).
The Kendall correlation coefficient is even more
 nonparametric than Spearman's correlation coefficient.
He uses only the relative ordering of ranks instead of using the 
numerical difference of ranks. Table~7 gives the results of our correlation
analyses.
 \begin{table}[htbp]
 \caption{Correlation coefficients (Pearson, Spearman, and  Kendall)
and probabilities for random correlations in percent for
continuum and  H$\beta$ line intensity variations versus
H$\beta$ line widths (FWHM) and H$\beta$ equivalent widths. }
    \centering
\tabcolsep+1.3mm
    \begin{tabular}[H]{lrrrrrr}
 \hline
  & $r_{p}$ & $r_{s}$  & $r_{k}$   & $P_{p}$&  $P_{s}$ & $P_{k}$ \\
 \hline
$F_{var}$  F$_{5100}$ vs FWHM    &  .637 &  .636 &  .511 & 4.8  & 5.9  & 4.0 \\     
$F_{var}$  H$\beta$   vs FWHM    &  .192 &  .179 &  .143 & 68.0 & 69.4 & 65.2 \\ 
$F_{var}$  F$_{5100}$ vs EQW     & -.119 & -.127 & -.111 & 74.4 & 68.9 & 65.5 \\ 
$F_{var}$  H$\beta$   vs EQW     &  .466 &  .500 &  .333 & 29.2 & 23.8 & 29.3 \\    
 \hline
 \end{tabular}
 \label{tab:stat}
 \end{table}
Those galaxies showing no significant variations in the
 H$\beta$ line intensity (Table~6)
were not considered for the H$\beta$ intensity correlation
analysis.

We found a significant correlation between the fractional variations of the 
continuum and the H$\beta$ line width in our very broad-line AGN sample.
The probability of being a random correlation is only 4 ($P_{k}$) or 6 
($P_{s}$) per cent. A correlation is said to be significant
if the probability for a random correlation is less than 5 per cent 
(Taylor J. R., \cite{taylor96}).

The fractional variations in the H$\beta$ line intensity are poorly correlated 
with the H$\beta$ equivalent width (Table~7).
No correlations were found between the the fractional variations in the
H$\beta$ line intensity and the H$\beta$ line widths,
as well as between the fractional variations
of the continuum and H$\beta$ equivalent widths. 
About one half of our AGN sample shows stronger variations of the continuum
than the other half (e.g. Fig.~\ref{fig:fvar_5100_vs_eqw_hb_o2}). 
The Spearman and
Kendall tests may fail to detect correlations in case of bimodality. 
However, both halves show the same relation between the fractional variations
of the continuum and H$\beta$ equivalent widths.  
Furthermore, Pearson's correlation coefficient gives the same results as
Spearman's and Kendall's tests.

\begin{figure}[htbp]
        \centering
        \includegraphics[width=8.5cm]{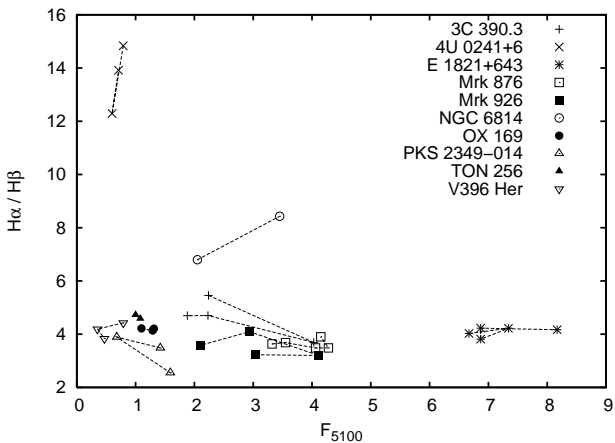}
        \caption{The Balmer decrement (H$\alpha$/H$\beta$)
          as a function of the continuum flux at 5100\,\AA{}. Errors are given
in Table~5.}
        \label{fig:balmerdec_vs_f5100}
\end{figure}

Figure~\ref{fig:balmerdec_vs_f5100} shows the Balmer decrement
(H$\alpha$/H$\beta$) and its variations
as a function of the continuum flux at 5100\,\AA{}.
Generally, the AGNs of our sample show Balmer decrements in the typical range 
of H$\alpha$/H$\beta$ $\simeq$3 to 5. This is consistent with little or 
moderate reddening ($\sim$2.8 to 3.1, case B). However, NGC\,6814 and 
4U0241$+$61 show pronounced Balmer decrements with 
H$\alpha$/H$\beta$ $\simeq$7 to 14, respectively. In both objects the broad 
H$\beta$ component is very weak (Figs.~\ref{fig:lk_1},\ref{fig:lk_2}). 
Considering the errors of the H$\alpha$ and the H$\beta$ intensities
(see Table~5), 
the Balmer decrement remained constant in most cases.
This result is independent
of absolute continuum and Balmer line fluxes. 
An expected anti-correlation of the Balmer decrement with the continuum 
flux variability cannot be confirmed by our sample.
Such variations in the Balmer decrement as a function of the ionizing 
continuum flux can be explained by radiative transfer effects rather 
than by variation in the dust extinction.
An increased optical depth in the Balmer lines can be induced by an
increasing ionizing flux. The ionization front penetrates deeper
into the gas in a radiation-bounded system.  Calculations by
Davidson \& Netzer (\cite{davidson79})
showed that the theoretical H$\alpha$/H$\beta$ ratio varies as 
a function of the optical depth. 
However, a nearly constant Balmer ratio 
in Fig~\ref{fig:balmerdec_vs_f5100} that is independent
of variations of the continuum
cannot be explained in a simple way. 
A complex behavior of the Balmer decrement as a function of continuum intensity
has been seen before, e.g.
in the variable Seyfert galaxy NGC~7603 (Kollatschny et al.
\cite{kollatschny00}).

An analysis of the fractional variations F$_{var}$ and the AGN luminosity, 
as well as redshifts, resulted in no significant 
correlation (see Sect. 4). 
\section{The variability analysis of further spectroscopic AGN samples}
The correlation between the
fractional continuum variations
and the H$\beta$ line widths in our AGN sample motivated us
to expand our sample and to test the relation 
on the published data from 
other campaigns investigating the optical variability of AGNs.

Kaspi et al. (\cite{kaspi00}) carried out an intense, long-term
monitoring project of PG quasars. Their spectra were taken at the
Wise Observatory and Steward Observatory.
They published the continuum intensities, H$\beta$ fluxes,
and H$\beta$ line widths (FWHM) of their  PG quasars.
The continuum and H$\beta$ fractional variability data of the PG
quasars (in Tables~8,9) were calculated by us from the original data
given by Kaspi et al. (\cite{kaspi00}). 
We determined the H$\beta$ equivalent line width from the H$\beta$ line 
fluxes and the continuum intensities.

\begin{table*}
\caption{AGN comparison sample selected from the variability campaigns of Kaspi
and Peterson.}
\tabcolsep+5.9mm
\begin{tabular}{lccccccc}
\hline
\noalign{\smallskip}
Object &  z & m$_{v}$ & M$_{B}$ & FWHM [OIII] &Ref.     &$\log{P_{5}}$ & Ref.   \\
&&& & kms$^{-1}$  &Col. 5   &W\,Hz$^{-1}$  & Col. 7  \\
(1) & (2) & (3) & (4)& (5)& (6) & (7) & (8)  \\  
\hline
 PG0026   &   0.142  &   15.41  &  -23.7   & 760   &1&22.59   &4 \\
 PG0052   &   0.155  &   15.43  &  -24.1   & 871   &1&21.83   &4 \\
 PG0804   &   0.100  &   14.71  &  -23.7   & 1513  &1&21.94   &4 \\
 PG0844   &   0.064  &   14.50  &  -22.1   & 1165  &1&20.66   &6 \\
 PG0953   &   0.239  &   15.32  &  -24.5   & 981   &1&23.45   &5 \\
 PG1211   &   0.085  &   14.19  &  -23.4   & 988   &1&23.61   &4 \\
 PG1226   &   0.158  &   12.85  &  -26.3   & 1582  &1&26.54   &4 \\
 PG1229   &   0.064  &   15.30  &  -21.7   & 827   &1&20.99   &4 \\
 PG1307   &   0.155  &   15.11  &  -23.5   & 845   &1&22.30   &5 \\
 PG1411   &   0.089  &   14.01  &  -22.9   & 1089  &1&21.25   &4 \\
 PG1426   &   0.086  &   14.87  &  -22.2   & 959   &1&21.51   &4 \\
 PG1613   &   0.129  &   15.49  &  -23.8   & 949   &1&22.28   &4 \\
 PG1617   &   0.114  &   15.39  &  -22.8   & 1107  &1&22.50   &5 \\
 PG1700   &   0.292  &   15.12  &  -25.2   & 1474  &1&24.40   &6 \\
 PG1704   &   0.371  &   15.28  &  -25.2   & 751   &1&25.80   &4 \\
 PG2130   &   0.061  &   14.64  &  -22.4   & 1044  &1&21.46   &4 \\
 3C 120   &   0.033  &   15.05  &  -20.8   & --    & &24.12   &4 \\
 3C 390.3 &   0.056  &   15.38  &  -21.6   & 799   &2&24.66   &4 \\
 Akn 120  &   0.032  &   13.92  &  -22.2   & 490   &3&22.43   &5 \\
 Fairall 9&   0.047  &   13.83  &  -23.0   & 425   &3&22.96   &5 \\
 IC 4329A &   0.016  &   13.66  &  -20.1   & 550   &3&22.26   &5 \\
 Mrk 110  &   0.035  &   15.37  &  -20.6   & 596   &2&21.24   &4 \\
 Mrk 279  &   0.030  &   14.46  &  -21.2   & 580   &3&22.15   &5 \\
 Mrk 335  &   0.026  &   13.85  &  -21.7   & 280   &3&21.60   &5 \\
 Mrk 509  &   0.034  &   13.12  &  -23.3   & 520   &3&21.27   &4 \\
 Mrk 590  &   0.026  &   13.81  &  -21.6   & 400   &3&22.40   &5 \\
 Mrk 79   &   0.022  &   14.27  &  -20.9   & 350   &3&22.07   &5 \\
 Mrk 817  &   0.031  &   13.79  &  -22.3   & 330   &3&22.03   &5 \\
 NGC 3783 &   0.010  &   14.43  &  -19.7   & 230   &3&21.63   &5 \\
 NGC 4051 &   0.002  &   12.92  &  -16.8   & 190   &3&20.60   &5 \\
 NGC 4151 &   0.003  &   11.85  &  -18.7   & 425   &3&21.79   &5 \\
 NGC 5548 &   0.017  &   13.73  &  -20.7   & 410   &3&21.91   &5 \\
 NGC 7469 &   0.016  &   13.04  &  -21.6   & 360   &3&22.58   &5 \\
\hline                                                               
\end{tabular}                              

Ref. 1: Calculated with data taken from
 \verb+http://wise-obs.tau.ac.il/~shai/PG/+ (Kaspi et al. \cite{kaspi00})\\
Ref. 2: Own calculations \\
Ref. 3: Peterson et al. \cite{peterson04}\\
Ref. 4: Xu et al. \cite{xu99}\\
Ref. 5: Ho \cite{ho02}\\
Ref. 6: Kellermann et al. \cite{kellermann89}\\
\end{table*}                              
\begin{table*}
\caption{Statistics of the continuum and H$\beta$ intensity variations,
H$\beta$ line widths, and equivalent widths
of AGN variability samples from Kaspi and Peterson.}
\tabcolsep+1.5mm
\begin{tabular}{lcccccccccccc}
\hline
\noalign{\smallskip}
Object    &  F$_{var}$  &  F$_{var}$  & Ref. &FWHM  H$\beta$& Ref.& EQW  H$\beta$ & Ref.
&  $F_{min}$ F$_{5100}$ & $F_{max}$ F$_{5100}$  & $F_{min}$ H$\beta$ & $F_{max}$ H$\beta$ & Ref.\\
          &  F$_{5100}$   &H$\beta$  &     &[kms$^{-1}$] & &[\AA{}] & \\
(1) & (2) & (3) & (4) & (5) & (6)& (7)&(8) & (9) & (10) & (11) & (12) & (13)\\
\hline
 PG0026   &   0.173  &  0.079  &1& 2100 $\pm{}$ 140 &2&   35  $\pm{}$ 4   &1     & 1.96 & 4.34  &  98.2  & 154.2 & 1 \\       
 PG0052   &   0.199  &  0.114  &1& 3990 $\pm{}$ 240 &2&   52  $\pm{}$ 10  &1     & 1.55 & 4.18  &  129.9 & 220.6 & 1 \\       
 PG0804   &   0.175  &  0.061  &1& 2984 $\pm{}$ 51  &2&   91  $\pm{}$ 15  &1     & 4.65 & 9.05  &  570.9 & 718.2 & 1 \\       
 PG0844   &   0.105  &  0.096  &1& 2730 $\pm{}$ 120 &2&   65  $\pm{}$ 4   &1     & 3.09 & 5.19  &  251.9 & 362.8 & 1 \\       
 PG0953   &   0.136  &  0.054  &1& 2885 $\pm{}$ 65  &2&   59  $\pm{}$ 8   &1     & 1.72 & 2.85  &  139.8 & 176.6 & 1 \\       
 PG1211   &   0.134  &  0.121  &1& 1832 $\pm{}$ 81  &2&   73  $\pm{}$ 7   &1     & 3.22 & 8.47  &  284.2 & 625.1 & 1 \\       
 PG1226   &   0.102  &  0.077  &1& 3416 $\pm{}$ 72  &2&   51  $\pm{}$ 6   &1     & 21.6 & 33.7  &  1408  & 1929  & 1 \\       
 PG1229   &   0.107  &  0.125  &1& 3440 $\pm{}$ 120 &2&   86  $\pm{}$ 8   &1     & 1.71 & 2.83  &  155.3 & 247.7 & 1 \\       
 PG1307   &   0.113  &  0.123  &1& 4190 $\pm{}$ 210 &2&   78  $\pm{}$ 9   &1     & 1.62 & 2.77  &  147.4 & 242.4 & 1 \\       
 PG1411   &   0.105  &  0.050  &1& 2456 $\pm{}$ 96  &2&   71  $\pm{}$ 5   &1     & 3.42 & 5.61  &  285.8 & 369.4 & 1 \\       
 PG1426   &   0.173  &  0.084  &1& 6250 $\pm{}$ 390 &2&   54  $\pm{}$ 6   &1     & 3.66 & 8.08  &  269.1 & 376.5 & 1 \\       
 PG1613   &   0.123  &  0.060  &1& 7000 $\pm{}$ 380 &2&   50  $\pm{}$ 6   &1     & 2.79 & 4.54  &  174.7 & 231.1 & 1 \\       
 PG1617   &   0.191  &  0.108  &1& 5120 $\pm{}$ 850 &2&   76  $\pm{}$ 17  &1     & 1.03 & 2.23  &  100.0 & 165.9 & 1 \\       
 PG1700   &   0.060  &  0.021  &1& 2180 $\pm{}$ 170 &2&   62  $\pm{}$ 4   &1     & 2.18 & 2.80  &  164.7 & 209.4 & 1 \\       
 PG1704   &   0.134  &  0.100  &1& 890  $\pm{}$ 280 &2&   11  $\pm{}$ 2   &1     & 1.56 & 2.88  &  23.6  & 39.7  & 1 \\       
 PG2130   &   0.086  &  0.089  &1& 2410 $\pm{}$ 150 &2&   95  $\pm{}$ 9   &1     & 4.19 & 6.36  &  400.1 & 595.9 & 1 \\       
 3C 120   &   0.180  &  0.095  &3& 2205 $\pm{}$ 185 &3 &   83  $\pm{}$ 13 &4     & 2.7  & 6.4  & 269.6  & 444.6   & 4\\      
 3C 390.3 &   0.343  &  0.072  &3& 9630 $\pm{}$ 804 &3 &   107 $\pm{}$ 15 &5     & 1.03 & 2.70 & 151.29 & 266.66  & 5\\       
 Akn 120  &   0.185  &  0.184  &3& 5284 $\pm{}$ 203 &3 &   90  $\pm{}$ 8  &4     & 6.6  & 12.7 & 570.8  & 1156.5  & 4\\       
 Fairall 9&   0.328  &  0.043  &3& 6901 $\pm{}$ 707 &3 &   98  $\pm{}$ 5  &5     & 5.31 & 7.88 & 409.16 & 631.31  & 5\\       
 IC 4329A &   0.115  &  0.084  &3& 6431 $\pm{}$ 6247&3 &                  &      & 3.41 & 4.54 & 288.4  & 384     & 9\\       
 Mrk 110  &   0.334  &  0.249  &3& 1521 $\pm{}$  59 &3 &   118 $\pm{}$ 24 &4     & 1.6  & 5.7  & 212.1  & 581.2   & 4\\       
 Mrk 279  &   0.092  &  0.058  &3& 3385 $\pm{}$ 349 &3 &   86  $\pm{}$ 8  &5     & 5.85 & 9.67 & 487.6  & 666.20  & 5\\       
 Mrk 335  &   0.090  &  0.060  &3& 1629 $\pm{}$ 145 &3 &   94  $\pm{}$ 7  &4     & 6.5  & 10.1 & 699.5  & 936.5   & 4\\       
 Mrk 509  &   0.181  &  0.106  &3& 2715 $\pm{}$ 101 &3 &   102 $\pm{}$ 15 &4     & 6.9  & 14.8 & 808.8  & 1466.2  & 4\\       
 Mrk 590  &   0.187  &  0.286  &3& 1979 $\pm{}$ 386 &3 &   46  $\pm{}$ 9  &4     & 4.7  & 10.2 & 138.5  & 605.3   & 4\\       
 Mrk 79   &   0.130  &  0.062  &3& 4219 $\pm{}$ 262 &3 &   74  $\pm{}$ 9  &4     & 5.2  & 10.0 & 467.2  & 671.2   & 4\\       
 Mrk 817  &   0.144  &  0.166  &3& 3515 $\pm{}$ 393 &3 &   70  $\pm{}$ 9  &4     & 4.2  & 8.0  & 282.6  & 514.4   & 4\\       
 NGC 3783 &   0.192  &  0.066  &3& 3093 $\pm{}$ 529 &3 &   82  $\pm{}$ 7  &5     & 9.68 & 14.68& 952    & 1273    & 5\\       
 NGC 4051 &   0.059  &  0.096  &3& 1072 $\pm{}$ 112 &3 &   44  $\pm{}$ 3  &5     & 10.75& 15.02& 370    & 589     & 5\\       
 NGC 4151 &   0.057  &  0.061  &3& 4248 $\pm{}$ 516 &3 &   102 $\pm{}$ 4  &5     & 72.20& 88.90& 6290   & 8430     & 5\\       
 NGC 5548 &   0.247  &  0.216  &3& 5421 $\pm{}$ 200 &3 &   85  $\pm{}$ 12 &5     & 4.89 & 17.46& 262    & 1206    & 5\\       
 NGC 7469 &   0.020  &  0.043  &3& 2169 $\pm{}$ 459 &3 &   51  $\pm{}$ 1  &5     &18.39 & 20.39& 709.41 & 853.8   & 5\\        
                                                                                                                                
 NGC 7603 &   0.427  &  0.538  &6& 6560 $\pm{}$ 100 &6 &   80  $\pm{}$ 17  &8    & 1.03 & 8.77 &  51    &  541   & 6\\        
 Mrk 110  &   0.318  &  0.189  &7& 1670 $\pm{}$ 50  &7 &   132 $\pm{}$ 26  &8    & 1.53 & 5.93 &  266   & 624.5  & 7\\       
\hline
\end{tabular}

Ref. 1: Calculated with data taken from
 \verb+http://wise-obs.tau.ac.il/~shai/PG/+ (Kaspi et al.  \cite{kaspi00})\\
Ref. 2: Kaspi et al.  (\cite{kaspi00})\\
Ref. 3: Peterson et al. (\cite{peterson04})\\
Ref. 4: Calculated with data taken from
 \verb+http://www.astronomy.ohio-state.edu/~peterson/AGN/+
(Peterson et al. \cite{peterson98})\\
Ref. 5: Calculated with data taken from
 \verb+http://www.astronomy.ohio-state.edu/~agnwatch/data.html+ \\
Ref. 6: Kollatschny et al. (\cite{kollatschny00})\\
Ref. 7: Bischoff et al. (\cite{bischkoll99})\\
Ref. 8: Own calculations \\
Ref. 9: Winge et al. (\cite{winge95})\\
\end{table*}
  \label{tab:petpgkaspidaten}
In big international projects such as the International AGN Watch campaign
(e.g. Alloin et al. \cite{alloin94}), the Lovers of Active Galaxies
program (e.g. Robinson \cite{robinson94}) and the Ohio State monitoring 
project (Peterson et al., \cite{peterson98})
about 30 AGNs have been investigated. 

Peterson et al. (\cite{peterson04}) present the results of a thorough
re-analysis of the entire AGN watch data,
employing improved methods.
From this paper we took
the fractional optical continuum and  
H$\beta$ line intensities variations, as well as the H$\beta$ line widths
of the AGNs and PG quasars (see Tables~8,9). We considered 
only data for those galaxies that were
monitored for periods of more than one year.
Furthermore we took only the average fractional variations
for individual galaxies over many years.
We measured the H$\beta$ equivalent widths of the AGNs in addition
to the PG quasars (Table~9), using the original AGN watch data
(Peterson et al. \cite{peterson04}),
or we calculated it based on the published H$\beta$ and continuum intensity
data (Peterson et al. \cite{peterson98}).

The fractional variation in the optical continuum of NGC~7469 
was taken from Peterson et al. (\cite{peterson98}).
Furthermore, the results of two other long-term monitoring projects
of NGC~7603 and Mrk~110 have been put in Table~9.
Spectroscopic results of  
these two AGNs have been published by Bischoff \& Kollatschny
(\cite{bischkoll99}) and Kollatschny et al. (\cite{kollatschny00}). 
The magnitudes M$_{B}$ and redshifts that we use in 
Figs.~\ref{fig:diplo2_over_diplP2_pete4_PGkaspi_hist_MB},
\ref{fig:diplo2_over_diplP2_pete4_PGkaspi_hist_z}
for the galaxies listed in Table~8, 9
have been taken from Veron \& Veron (\cite{Veron01}). 
\subsection{Combined variability sample}
\begin{figure}[htbp]
        \centering
        \includegraphics[width=8.5cm]{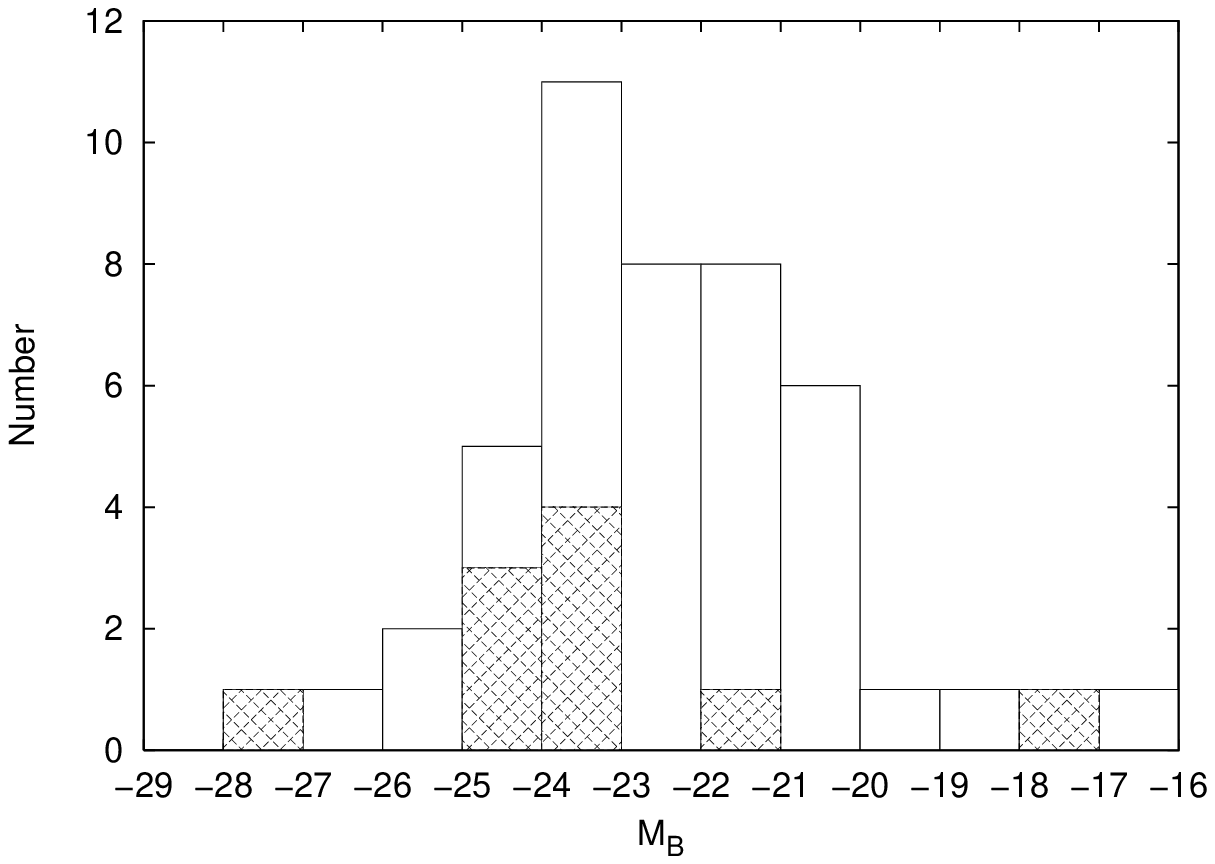}
        \caption{Distribution of absolute B magnitudes of the combined AGN
       sample.
   The hatched area gives the distribution of the very broad-line AGN sample}
        \label{fig:diplo2_over_diplP2_pete4_PGkaspi_hist_MB}
\end{figure}

\begin{figure}[htbp]
        \centering
        \includegraphics[width=8.5cm]{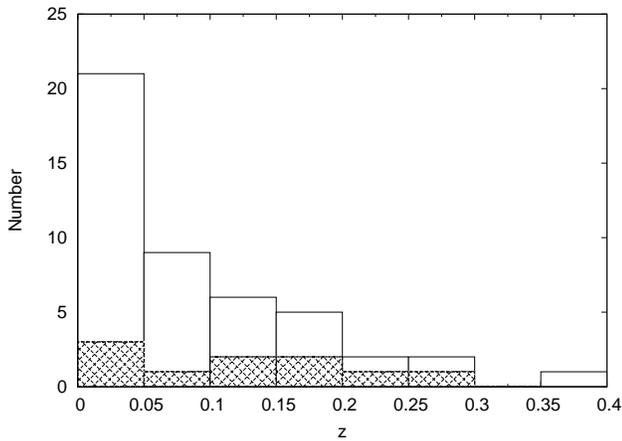}
        \caption{Redshift distribution of the combined AGN sample.
   The hatched area gives the distribution of the very broad-line AGN sample.}
        \label{fig:diplo2_over_diplP2_pete4_PGkaspi_hist_z}
\end{figure}

We combined the data for our variability sample of very broad-line AGNs
with other AGN data taken from the literature (Table~8).
Figure~\ref{fig:diplo2_over_diplP2_pete4_PGkaspi_hist_MB}
shows the distribution of absolute B magnitudes
of the combined AGN sample. 
In Fig.~\ref{fig:diplo2_over_diplP2_pete4_PGkaspi_hist_z}
we plot the redshift distribution of the combined AGN sample.
\begin{figure}[htbp]
        \centering
        \includegraphics[width=8.5cm]{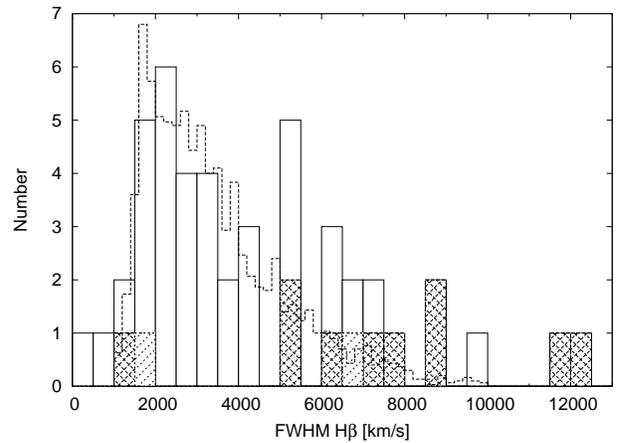}
        \caption{H$\beta$ line widths distribution (FWHM) of the
          combined AGN sample in steps 500 kms$^{-1}$. The hatched area
          gives the distribution of the very broad-line AGN sample. 
          The dashed line gives the scaled-down distribution 
         of  H$\alpha$ line widths  of AGNs derived from the 
         Sloan Digital Sky Survey (Hao et al. \cite{hao05}).}
        \label{fig:hao_over_diplo2_over_diplP2_pete4_PGkaspi_hist_hwb_hb}
\end{figure}
\begin{figure}[htbp]
        \centering
        \includegraphics[width=8.5cm]{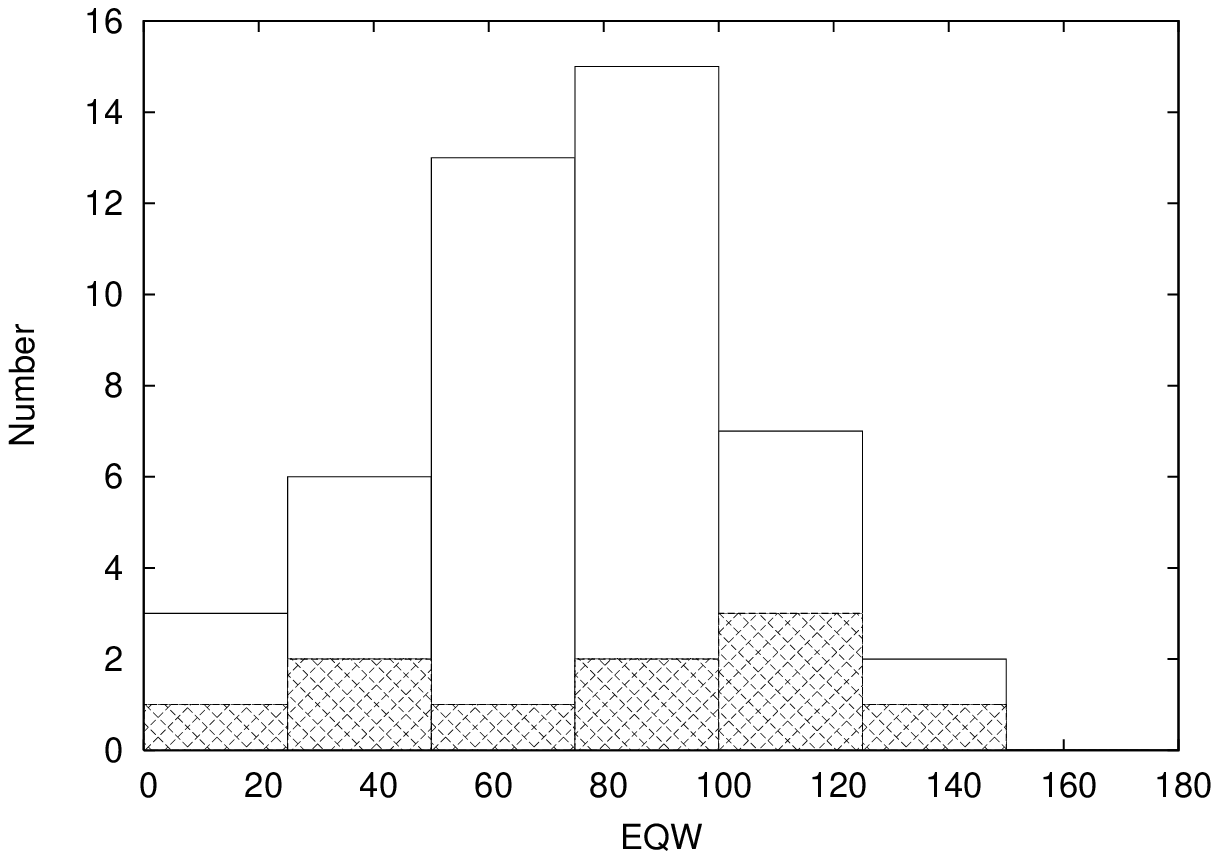}
        \caption{H$\beta$ equivalent width distribution of the combined AGN
      sample. The hatched area gives the distribution of the very broad
         line AGN sample.}
        \label{fig:diplo2_over_diplP2_pete4_PGkaspi_hist_eqw}
\end{figure}

We show the H$\beta$ line widths distribution (FWHM) of the
combined AGN sample
in Fig.~\ref{fig:hao_over_diplo2_over_diplP2_pete4_PGkaspi_hist_hwb_hb}.
We used a step-size of 500 kms$^{-1}$. 
The cross-hatched bars give the distribution
of the very broad-line AGN sample. 
The two single-hatched bars are based on
G\"ottingen long-term variability studies of the galaxies Mrk~110 and NGC~7603.
The relative distribution of the H$\alpha$ line widths (FWHM) of AGNs derived 
from the Sloan Digital Sky Survey (SDSS) (Hao et al. \cite{hao05})
is superposed in 
Fig.~\ref{fig:hao_over_diplo2_over_diplP2_pete4_PGkaspi_hist_hwb_hb}. 
The H$\beta$ FWHM distribution of our combined variability AGN sample 
is representative of the H$\alpha$ FWHM distribution of all SDSS AGNs,
but with an excess of very broad-lined objects.
Figure~\ref{fig:diplo2_over_diplP2_pete4_PGkaspi_hist_eqw}
then gives the H$\beta$ equivalent width distribution of the
very broad-line AGN sample and of the comparison sample.

In general, the distributions of the H$\beta$ equivalent line widths,
the absolute galaxy magnitudes, and the redshifts in the very broad-line
AGN sample and the comparison sample are very similar. Since
very broad-line AGNs are rare objects, their spectra had to be taken at
slightly larger redshifts on average.
 Since the apparent magnitudes of the galaxies
are similar in both variability samples,
the absolute magnitudes of the very broad-line objects are therefore
slightly brighter.
\subsection{Correlation results of the combined sample}
\begin{figure}[htbp]
        \centering
        \includegraphics[width=8.5cm]{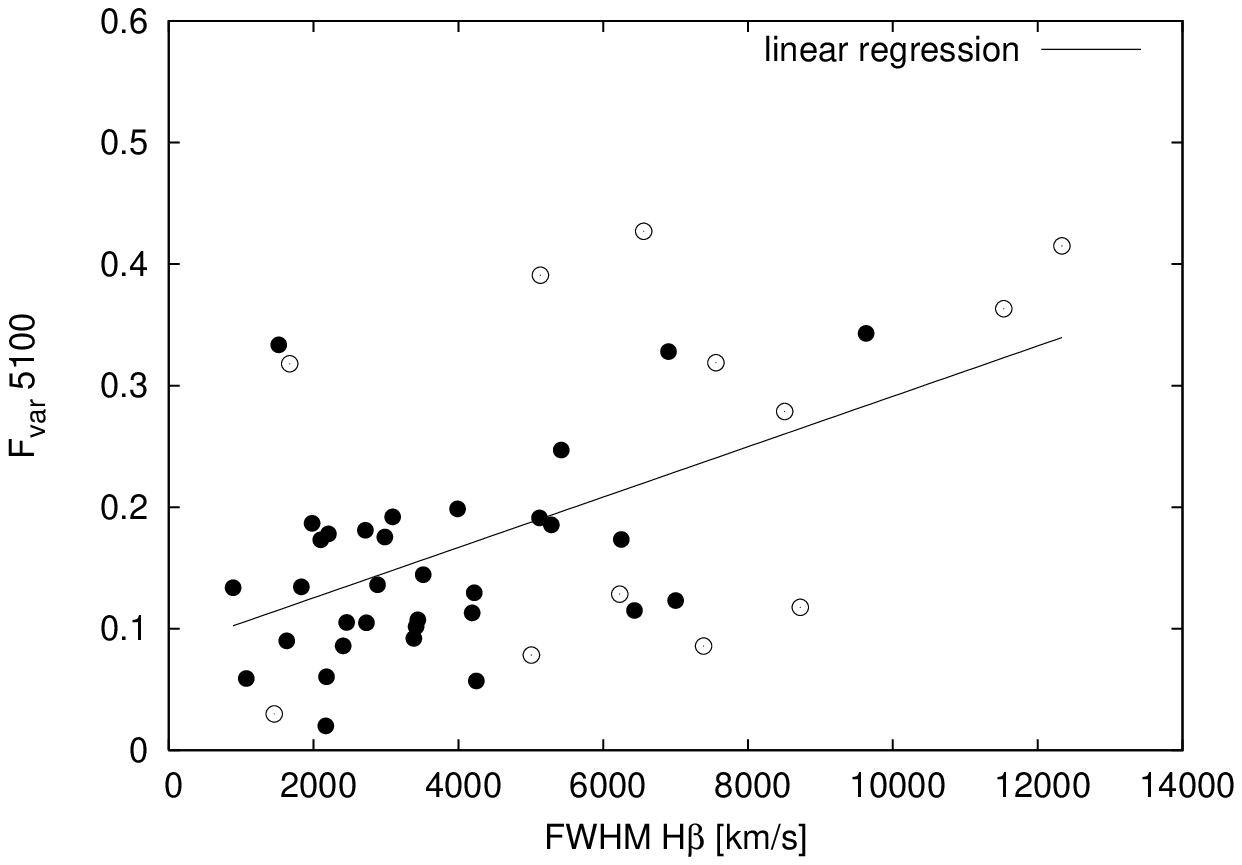}
        \caption{Fractional variation of the continuum at 5100\,\AA\
 versus H$\beta$ FWHM (combined AGN sample). Shown are the
very broad-line AGN sample, NGC 7603, and Mrk~110 (open circles),
        and Kaspi and Peterson comparison sample (filled circles).}
        \label{fig:diplP2_pete4_PGkaspi_fvar_5100_vs_hwb_hb}
\end{figure}

\begin{figure}[htbp]
        \centering
        \includegraphics[width=8.5cm]{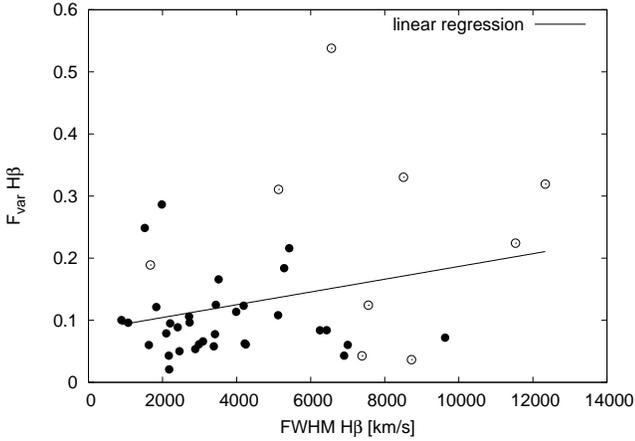}
        \caption{Fractional variation in H$\beta$
          versus H$\beta$ FWHM (combined AGN sample).}
        \label{fig:diplP2_pete4_PGkaspi_fvar_hb_vs_hwb_hb}
\end{figure}

\begin{figure}[htbp]
        \centering
        \includegraphics[width=8.5cm]{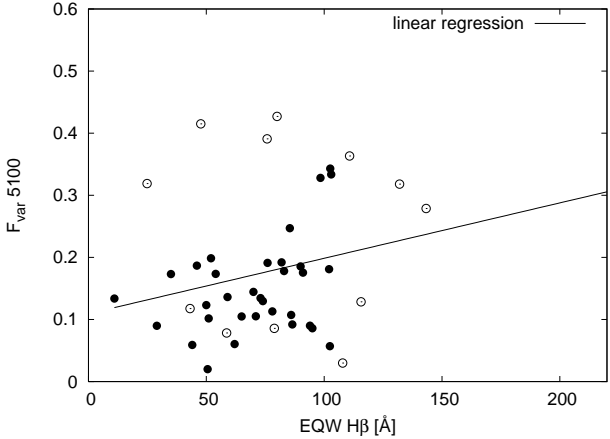}
        \caption{Fractional variation of the continuum at 5100\,\AA\
          versus H$\beta$ equivalent width (combined AGN sample).}
        \label{fig:diplP2_pete4_PGkaspi_fvar_5100_vs_eqw_hb}
\end{figure}

\begin{figure}[htbp]
        \centering
        \includegraphics[width=8.5cm]{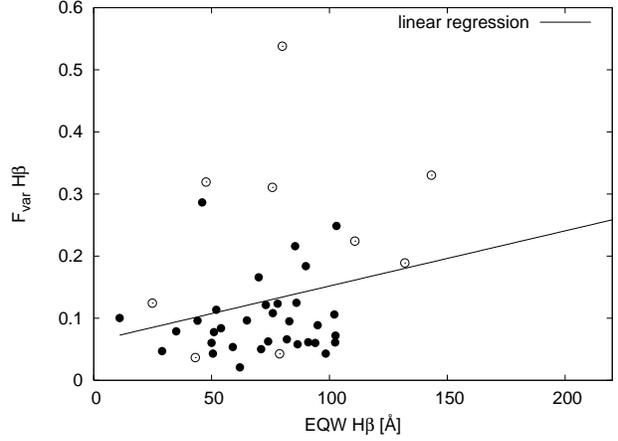}
        \caption{Fractional variation in H$\beta$
          versus H$\beta$ equivalent width (combined AGN sample).}
        \label{fig:diplP2_pete4_PGkaspi_fvar_hb_vs_eqw_hb}
\end{figure}

\begin{figure}[htbp]
        \centering
        \includegraphics[width=8.5cm]{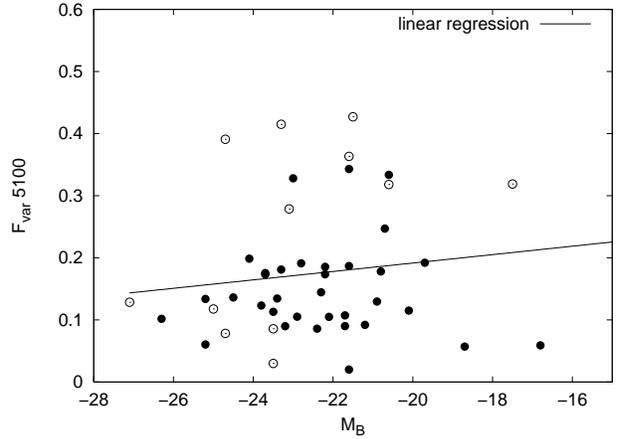}
        \caption{Fractional variation of the continuum at 5100\,\AA\
          versus the blue magnitude of the AGNs (combined AGN sample).}
        \label{fig:diplP2_pete4_PGkaspi_fvar_5100_vs_MB}
\end{figure}

\begin{figure}[htbp]
        \centering
        \includegraphics[width=8.5cm]{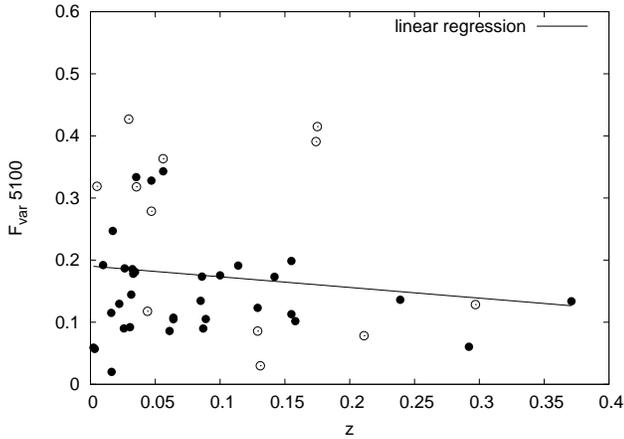}
        \caption{Fractional variation of the continuum at 5100\,\AA\
          versus redshift (combined AGN sample).}
        \label{fig:diplP2_pete4_PGkaspi_fvar_5100_vs_z}
\end{figure}

\begin{figure}[htbp]
        \centering
        \includegraphics[width=8.5cm]{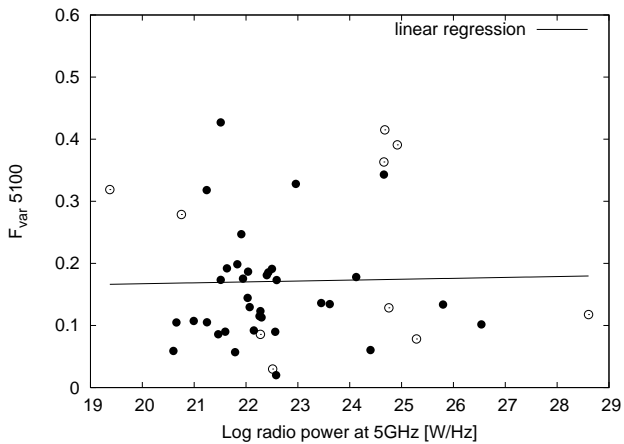}
        \caption{Fractional variation of the continuum at 5100\,\AA\
          versus radio power at 5~Ghz (combined AGN sample).}
        \label{fig:diplP2_pete4_PGkaspi_fvar_5100_vs_radio_gemittelt}
\end{figure}

\begin{figure}[htbp]
        \centering
        \includegraphics[width=8.5cm]{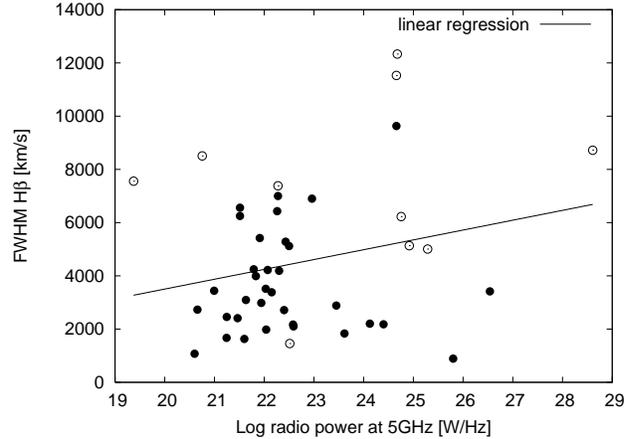}

        \caption{H$\beta$ line width (FWHM)
          versus radio power at 5~Ghz (combined AGN sample).}
        \label{fig:diplP2_pete4_PGkaspi_hwb_hb_vs_radio_gemittelt}
\end{figure}

The line and continuum data of the combined AGN sample are given 
in Tables 6~and~9. The combined sample consists
of 43 different galaxies.
The galaxies Mrk~110 and 3C390.3 have been observed both by us
(very broad-line AGN sample)
and by other groups (see text). Therefore, their values are 
shown twice in  Figs.~\ref{fig:diplP2_pete4_PGkaspi_fvar_5100_vs_hwb_hb} to 
\ref{fig:diplP2_pete4_PGkaspi_hwb_hb_vs_radio_gemittelt}. 
We computed a mean of their observed values which are 
used in the correlation analysis. The resulting correlation coefficients are 
given in Tables 10~and~11. 
There we consider only one value for each galaxy.

The optical continuum and  H$\beta$ line intensity variations are plotted over 
the H$\beta$ line widths  and their equivalent widths in 
Figs.~\ref{fig:diplP2_pete4_PGkaspi_fvar_5100_vs_hwb_hb} to 
\ref{fig:diplP2_pete4_PGkaspi_fvar_hb_vs_eqw_hb} for 
the combined sample. Again we show
the linear fit derived from the Pearson correlation
for comparison.
Table~10 gives the results of our correlation analyses of the combined sample.
 \begin{table}[htbp]
 \caption{Correlation coefficients (Pearson, Spearman, and Kendall)
and probabilities for random correlations in percent for
continuum and  H$\beta$ line intensity variations versus
H$\beta$ line widths (FWHM) and H$\beta$ equivalent widths
(combined AGN sample)}
    \centering
    \tabcolsep+1.3mm
    \begin{tabular}[H]{lrrrrrr}
 \hline
  & $r_{p}$ & $r_{s}$  & $r_{k}$   & $P_{p}$&  $P_{s}$ & $P_{k}$ \\
 \hline
$F_{var}$  F$_{5100}$ vs FWHM   & .553 & .409 & .300 &  0.0 &  0.8 &  0.5 \\
$F_{var}$  H$\beta$   vs FWHM   & .322 & .157 & .114 &  4.2 & 32.7 & 29.9 \\
$F_{var}$  F$_{5100}$ vs EQW    & .156 & .125 & .079 & 31.8 & 41.6 & 45.7 \\ 
$F_{var}$  H$\beta$   vs EQW    & .214 & .170 & .122 & 18.4 & 28.9 & 26.8 \\
 \hline
 \end{tabular}
\label{tab:stat2}
\end{table}
The combined AGN sample confirms and strengthens the correlation results
we got for the very broad-line sample just above: \\
- No correlation was found between the fractional variations
  of the continuum and H$\beta$ equivalent widths.
  A correlation mentioned by Giveon (\cite{giveon99}) for his sample of
  Palomar-Green quasars cannot be confirmed by our results.
  The H$\beta$ equivalent widths of  Boroson and Green (\cite{boroson92})
 used by
  Giveon et. al (\cite{giveon99}) were  taken one year before the start of
 their variability campaign. \\
- The fractional variations in the H$\beta$ line intensities
   are poorly correlated with the H$\beta$
 equivalent widths, as well as with the H$\beta$ line widths.
  Pearson's test gives a reasonably strong correlation for H$\beta$ variation
  vs FWHM H$\beta$, whereas the other tests do not. However, 
  Pearson's correlation coefficient is a parametric statistic, and it may be
  less useful if the underlying assumption of normality is violated.\\
- But a significant correlation resulted for the fractional variations of the 
  optical continua with the H$\beta$ line widths. The probability of being a
  random correlation is 0.5 ($P_{k}$) only or 0.8 ($P_{s}$) per cent.

Next, we studied the relation of F$_{var}$(5100) and F$_{var}$(H$\beta$)
with redshift and luminosity for possible evolutionary effects (Table~11),
and found no correlation between the fractional continuum variation with
luminosities
(see Fig.~\ref{fig:diplP2_pete4_PGkaspi_fvar_5100_vs_MB}).
There is no correlation between
the fractional variations of the continuum in the rest-frame at 5100 \AA\
and the AGN redshifts 
(72 ($P_{k}$) or 75 ($P_{s}$) per cent)(Fig.~\ref{fig:diplP2_pete4_PGkaspi_fvar_5100_vs_z}) . 

 \begin{table}[htbp]
 \caption{Correlation coefficients (Pearson, Spearman, and Kendall)
and probabilities for random correlations in percent for
various AGN properties.
For continuum and  H$\beta$ line intensity variations versus
radio power at 5~Ghz,  
H$\beta$  equivalent widths, and H$\beta$ line widths (FWHM)
versus radio power at 5~Ghz,
continuum variations versus AGN magnitudes, and versus
redshifts (combined AGN sample),
as well as black-hole mass and black-hole mass/luminosity ratio
 versus H$\beta$ line widths (M$_{BH}$ in units of $10^{6}$ M$_{\odot}$, 
 L$_{44}$: optical luminosity in units of $10^{44}$ erg s$^{-1}$).}
    \centering
    \tabcolsep+1.mm
    \begin{tabular}[H]{lrrrrrr}
 \hline
  & $r_{p}$ & $r_{s}$  & $r_{k}$   & $P_{p}$&  $P_{s}$ & $P_{k}$ \\
 \hline
$F_{var}$ F$_{5100}$ vs $\log{P_{5}}$ &  .018 &  .025 &  .034 & 90.7 & 86.9 & 74.6 \\
$F_{var}$  H$\beta$  vs $\log{P_{5}}$ & -.100 & -.211 & -.155 & 53.3 & 18.1 & 15.3 \\
H$\beta$ EQW         vs $\log{P_{5}}$ & -.217 & -.198 & -.139 & 16.2 & 19.9 & 19.0 \\
H$\beta$ FWHM        vs $\log{P_{5}}$ &  .205 &  .090 &  .061 & 18.7 & 56.2 & 56.4 \\
$F_{var}$ F$_{5100}$ vs M$_{B}$     &  .091 &  .117 &  .096 & 55.5 & 44.3 & 36.0 \\  
$F_{var}$ F$_{5100}$ vs z           & -.108 & -.049 & -.037 & 48.6 & 74.8 & 72.3 \\
M$_{BH}$ vs FWHM H$\beta$           &  .191 &  .239 &  .173 & 29.4 & 18.4 & 16.3 \\
M$_{BH}$/L$_{44}$ vs FWHM H$\beta$  &  .509 &  .398 &  .294 &  0.3 &  2.7 &  1.8 \\   
\hline
 \end{tabular}
\label{tab:stat1}
\end{table}
We also tested whether there are any dependencies in the behavior of the
targets as a function of their radio power.
We found  no correlation between the continuum
or H$\beta$ line intensity variations 
with the radio power at 5~Ghz
(Table~11). Figure~\ref{fig:diplP2_pete4_PGkaspi_fvar_5100_vs_radio_gemittelt}
 e.g. shows the fractional variation of the continuum
at 5100\,\AA\ versus radio power at 5~Ghz.\\  
Furthermore, there is no significant correlation between the
H$\beta$ line widths (FWHM) or H$\beta$  equivalent widths 
with the radio power at 5~Ghz (Table~11). 
Figure~\ref{fig:diplP2_pete4_PGkaspi_hwb_hb_vs_radio_gemittelt}
e.g. shows the H$\beta$ line width (FWHM) versus radio power at 5~Ghz.
\section{Discussion and conclusion}
First we discuss probable luminosity
or redshift correlations
before going into detail regarding the correlations of
the continuum variations with the line-widths. 

\subsection{Redshift -- variability correlations}
So far, most of the information on AGN variability has been 
derived only from light curves in single bands.
An average increase in variability with redshift of galaxies was announced
by e.g. Cristiani et al. (\cite{christiani96}).
 Trevese \& Vagnetti (\cite{Trevese02}) reanalyzed
 B and R light curves of PG quasars. They confirm the correlation between  
variability and redshift. But because B and R light curves
correspond to different intrinsic wavelength bands in the quasar spectra, 
they explain the increase in the amplitudes with
the increase in rest-frame frequency at higher redshifts.
It is known that AGNs vary more in the blue  than in the red. 
De Vries et al. (\cite{devries05}) for instance investigated
the long-term quasar variability
from Sloan Digital Sky Survey (SDSS) data, to find that
the magnitude of the quasar variability is a function of wavelength.
The variability increases toward the blue part of the spectrum.

In our present investigation, we derived the continuum variability amplitude
from the AGN spectra in the rest-frame at 5100\AA{}. 
There is no correlation between redshift
and rest-frame continuum variability amplitude (Fig.~\ref{fig:diplP2_pete4_PGkaspi_fvar_5100_vs_z}).
Only 5 galaxies in our sample have  redshifts higher than 0.2.
It would be useful to examine the spectral variability of other objects
at larger distances.

\subsection{Luminosity -- variability correlations}
Giveon et al. (\cite{giveon99}) found trends of an anti-correlation
of variability amplitude with luminosity.  De Vries et al. (\cite{devries05}) 
confirmed this trend in their long-term quasar variability study, showing that 
high-luminosity quasars vary less than low-luminosity quasars.
They explain it
with a scenario in which variations in AGNs
caused by chromatic outbursts or flares
have a limited absolute magnitude.
In our variability study we also find the basic trend that more luminous 
objects are less variable
(see Fig.~\ref{fig:diplP2_pete4_PGkaspi_fvar_5100_vs_MB}), 
but this trend is not highly significant.

The fractional variability of the continuum of our very broad-line AGN sample 
shows either small or large variations.
This suggests a bimodality for the very broad-line AGN sample.
However, the bimodality of the fractional variations of the continuum 
disappears in the combined AGN sample.

\subsection{X-ray -- linewidth correlations}
Many studies deal with correlations between different 
AGN properties. But not many detailed investigations exist
that look for correlations
between optical continuum variations and optical line widths.
Fiore et al. (\cite{fiore98}) studied the X-ray variability properties of six
PG quasars. They correlated the X-ray variability properties with optical 
line widths and found evidence that the three narrow-line AGNs of their
sample show larger X-ray variability amplitudes
than the three broader-line AGNs on timescales of 2 to 20 days.
No differences were found on longer timescales. 
They propose that their correlation between X-ray amplitude variations
and narrow/broad-line AGNs were caused by different L/$L_{Edd}$ states,
and suggest that narrow-line, steep X-ray
spectrum AGNs emit close to the Eddington luminosity and
have a relatively low-mass black hole. 

We found no correlation between black hole masses and
optical line widths (Table~11) or variability amplitudes (see next section). 
Our study is concentrated on long-term variations over years,
on the one hand, and 
variations in the optical, on the other,
contrary to the study of Fiore et al. (\cite{fiore98}).
Variations in different frequency ranges may have completely different
origins.

\subsection{Optical variability -- linewidth correlations}

First of all, we tested whether there is a correlation between
continuum variations and H$\beta$ equivalent widths. 
Giveon et al. (\cite{giveon99})
compared the long-term optical variability properties of PG quasars
with many parameters.
They found trends to an increase in continuum
variability amplitudes with H$\beta$ equivalent width.
As a possible, but unlikely, explanation for their trend,
they discussed
the possibility of different contributions of the
 emission lines to their variable broad-band fluxes.
We tested the variability amplitude and H$\beta$ equivalent width
 correlation again for our samples of very broad-line
and normal AGNs.
In our study,
the continuum flux was taken from the individual spectra and not
from broad-band data.
Our continuum flux is therefore independent
of emission line contribution.
We find no indication of a correlation between the 
continuum and H$\beta$ variability amplitudes with the H$\beta$ equivalent 
width.

But there is a
significant correlation between AGN continuum variability amplitudes
and H$\beta$ line widths
for our combined AGN sample (see Table~7 and
Fig.~\ref{fig:diplP2_pete4_PGkaspi_fvar_5100_vs_hwb_hb}). 
This confirms the earlier result of our very broad-line sample. 
A relationship between these parameters has
not been considered earlier by Giveon et al. (\cite{giveon99}).

The observed line width of the broad emission lines (including  H$\beta$)
in AGN spectra might be caused or
affected by different parameters simultaneously:\\
- Different central black-hole masses lead to different
Keplerian velocities of gravitationally bound emission-line clouds.\\
- The emission lines might originate at different radii with respect
to identical central black hole masses, again making
the assumption that their velocities are Keplerian.\\
- If the emission line region is connected with a central accretion disk,
the observed line width might be correlated with the inclination 
of the disk (Kollatschny \cite{kollatschny03a}) to our line-of-sight.\\
- Optical depth effects on line profiles produced in an accretion disk
wind (Murray \& Chiang \cite{murray97}) might affect the profiles.
It has been shown that the observed single-peaked profiles in Mrk~110 are
consistent enough to be
generated in an accretion  disk (Kollatschny \& Bischoff \cite{kollatschny02})
indicating optical depth effects.
Optically thin lines generated in an accretion disk
 should have double-peaked profiles.

The observed line-width in AGNs might be caused by a combination of these
different effects in practice. 
The importance of the individual
parameters could be different from galaxy to galaxy.
Therefore, the strong significance of 
 the correlation between  H$\beta$ line width and
continuum variations is surprising. 

We tested whether the central black-hole mass is correlated with the
width of the broad H$\beta$ line. By taking the black-hole mass data from
the Peterson '04 sample (\cite{peterson04}) and correlating them with
the H$\beta$ line widths, we found no correlation
(see Table~11).
But  the widths of the stellar absorption lines seem
to be correlated with the central black-hole mass (Onken et al.
\cite{onken04}, Ferrarese \& Merritt \cite{ferrarese00}).

\begin{figure}[htbp]
        \centering
        \includegraphics[width=9cm]{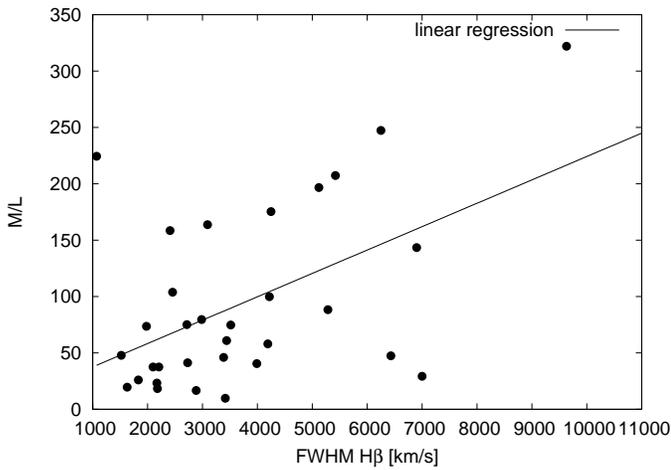}
        \caption{M/L vs. H$\beta$ line width (for Peterson '04 sample).}
        \label{fig:fvar_5100_vs_MdL}
\end{figure}

Furthermore, we tested the significance of the correlation between
H$\beta$ line
width  and black-hole mass-to-luminosity ratio. 
Figure~\ref{fig:fvar_5100_vs_MdL} shows this relation and
Table~11 gives their significance.
Wandel (\cite{wandel02}) has shown earlier that there is a significant
correlation between these two parameters. He explains the close 
correlation between mass-to-luminosity ratio and line-width
with the virial relation $M_{BH} \propto
v(FWHM)^{2}$.
But on the basis of the virial relation alone, 
it is difficult to understand
 why the correlation
$M_{BH}/L \propto v(FWHM)^{2}$  is far more significant than that of
$M_{BH} \propto v(FWHM)^{2}$.

Comparing the significance of the correlation between  
black-hole mass-to-luminosity ratio and line width
as a random correlation ($P_{k}$ = 1.8 percent, Table~11)
with our earlier correlation between 
optical continuum
variability amplitudes and  H$\beta$ emission line widths
as a random correlation ($P_{k}$ = 0.5 percent, Table~10),
this relation is equal or even better.

Considering the influence of the inclination angle on
broad emission-line widths
there are indications that
the line width of the broad emission lines in AGNs
 depends on the inclination angle of a central accretion disk 
(Wills \& Browne \cite{wills86},
Kollatschny \cite{kollatschny03b}). But it is not easy to understand why
the continuum variability of an edge on AGN should be stronger
than the continuum variability of a face-on central source.
Optical thickness effects might be important (Netzer, \cite{netzer87}). 

Considering individual galaxies, 
it is known that broader emission lines originate closer
to the center and show stronger
variability amplitudes (e.g. Kollatschny et al.
\cite{kollatschny01}). 
The central ionizing sources in 
AGNs - showing extreme flux variations over timescales of years -
might excite BLR clouds over wider ranges
in an accretion disk (and therefore produce 
lines with different widths)
than those AGNs showing small-scale flux variations only. 

In summary, our variability study showed that the most
significant predictor of optical
variability properties is the emission-line width. But the close
correlation between these two parameters is not easy to explain.

\begin{acknowledgements}
      Part of this work was supported by the
      \emph{Deut\-sche For\-schungs\-ge\-mein\-schaft, DFG\/}. 
\end{acknowledgements}
\end{document}